\documentclass[12pt,letterpaper]{article}

\usepackage{etoolbox}
\newtoggle{SUPPLEMENTAL}\toggletrue{SUPPLEMENTAL}
\togglefalse{SUPPLEMENTAL} 
\newcommand{\Supplemental}[2]{\iftoggle{SUPPLEMENTAL}{#1}{#2}}
\newtoggle{BLINDED}\toggletrue{BLINDED}
\togglefalse{BLINDED} 
\newcommand{\Blinded}[2]{\iftoggle{BLINDED}{#1}{#2}}

\usepackage{graphicx,grffile}
\usepackage{amsmath,amssymb,amsthm}
\usepackage{mathtools,dsfont,centernot}
\usepackage{caption}
\usepackage{booktabs,tabularx,threeparttable,longtable,ragged2e}
\usepackage{siunitx}
\usepackage[shortlabels]{enumitem}
\usepackage{alltt}
\usepackage[longnamesfirst]{natbib}
\usepackage{hypernat}
\usepackage[nodisplayskipstretch]{setspace}
\usepackage[bottom]{footmisc}
\usepackage{xr} 
\usepackage{xcolor}

\usepackage[hyphens]{url}
\usepackage{hyperref}

\usepackage[margin=1in,letterpaper]{geometry}
\usepackage[capitalize,noabbrev]{cleveref}

\usepackage{scalerel}
\newcommand\varint[1][.8]{\mathrel{\hstretch{#1}{\int}}}

\usepackage{tikz}
\usetikzlibrary{shapes,calc}
\usetikzlibrary{patterns}

\usepackage{algorithm,algorithmic}

  \renewenvironment{thebibliography}[1]%
  {\begin{oldthebibliography}{#1}\setlength{\parskip}{0ex}\setlength{\itemsep}{0ex}}%
  {\end{oldthebibliography}}
\appto\TPTnoteSettings{\linespread{1}\footnotesize}

\DeclareGraphicsExtensions{.pdf,.png}
\graphicspath{ {./Figures/} }

\usepackage[page]{appendix}
\renewcommand{\appendixpagename}{\LARGE \centering Online Appendix for ``A quantile-based nonadditive fixed effects model''} 

\usepackage{lmodern}
\usepackage{adjustbox}

\newcommand{\citeposs}[1]{\citeauthor{#1}'s (\citeyear{#1})}

\crefname{conjecture}{Conjecture}{Conjectures}
\crefname{section}{Section}{Sections}
\crefname{subsection}{Section}{Sections}
\crefname{subsubsection}{Section}{Sections}
\Crefname{conjecture}{Conjecture}{Conjectures}
\Crefname{section}{Section}{Sections}
\Crefname{subsection}{Section}{Sections}
\Crefname{subsubsection}{Section}{Sections}
\crefname{appendix}{Appendix}{Appendices}
\crefname{subappendix}{Appendix}{Appendices}
\crefname{subsubappendix}{Appendix}{Appendices}
\Crefname{appendix}{Appendix}{Appendices}
\Crefname{subappendix}{Appendix}{Appendices}
\Crefname{subsubappendix}{Appendix}{Appendices}
\crefname{equation}{}{}
\Crefname{equation}{Equation}{Equations}
\crefname{enumi}{}{}
\Crefname{enumi}{}{}

\crefname{assumption}{}{}
\Crefname{assumption}{Assumption}{Assumptions}

\Crefname{method}{Method}{Methods}
\newlist{steps}{enumerate}{1}
\setlist[steps]{label=Step \arabic*:, ref=\arabic*, itemsep=0pt}
\crefname{stepsi}{Step}{Steps}
\Crefname{stepsi}{Step}{Steps}

\theoremstyle{plain}
\newtheorem{theorem}{Theorem}
\newtheorem{lemma}[theorem]{Lemma}
\newtheorem{proposition}[theorem]{Proposition}
\newtheorem{corollary}[theorem]{Corollary}

\theoremstyle{definition}

\newtheorem{assumption}{Assumption}

\newcommand{\numnornd}[1]{\num[round-mode=none,group-digits=integer]{#1}} 
\newcommand{\ubar}[1]{\mkern3mu\underline{\mkern-3mu #1\mkern-3mu}\mkern3mu}
\newcommand{\matf}[1]{\ubar{\boldsymbol{\mathbf{#1}}}} 
\newcommand{\vecf}[1]{\boldsymbol{\mathbf{#1}}} 

\newcommand{\iid}{\stackrel{\mathit{iid}}{\sim}}
\newcommand{\asconv}{\xrightarrow{a.s.}}

\newcommand{\pconv}{\xrightarrow{p}}
\newcommand{\dconv}{\xrightarrow{d}}

\DeclareMathOperator{\Cov}{Cov}

\newcommand{\R}{{\mathbb R}}

\DeclareMathOperator{\E}{E} 
\DeclareMathOperator{\Q}{Q} 
\let\Pr\relax \DeclareMathOperator{\Pr}{P} 

\DeclareMathOperator{\1}{\mathds{1}}
\newcommand{\Ind}[1]{\1\{#1\}}

\newcommand{\BetaDist}{\textrm{Beta}}
\newcommand{\DirDist}{\textrm{Dir}}
\newcommand{\UnifDist}{\textrm{Unif}}

\newcommand{\weaklyto}{\rightsquigarrow}

\newcommand{\independenT}[2]{\mathrel{\rlap{$#1#2$}\mkern2mu{#1#2}}}
\newcommand\independent{\protect\mathpalette{\protect\independenT}{\perp}} 
\providecommand{\abs}[1]{\lvert#1\rvert}
\providecommand{\absbig}[1]{\left\lvert#1\right\rvert}
\providecommand{\norm}[1]{\lVert#1\rVert}
\providecommand{\normbig}[1]{\left\lVert#1\right\rVert}

\let\originalleft\left
\let\originalright\right
\renewcommand{\left}{\mathopen{}\mathclose\bgroup\originalleft}
\renewcommand{\right}{\aftergroup\egroup\originalright}

\newcommand{\mockalph}[1]{}  
\allowdisplaybreaks[3]

\usepackage{dcolumn}
\newcolumntype{d}[1]{D{.}{\cdot}{#1} }

\hypersetup{
  pdfauthor = {Xin Liu},
  pdfkeywords = {economics, econometrics, statistics, panel data},
  pdftitle = {quantile-based nonadditive FE},
  pdfsubject = {econometrics},
  pdfpagemode = UseNone
}

\defcitealias{ChernozhukovHansen2005}{CH05}

\title{A quantile-based nonadditive fixed effects model}

\author{\Blinded{[BLINDED]}{%
Xin Liu%
\thanks{
        School of Economic Sciences, Washington State University.
        Email: \texttt{xin.liu1@wsu.edu}.
}
}}

\date{\today}

\begin{document}

\maketitle

\justifying

\begin{abstract}
\newcommand{\paperspacing}{\onehalfspacing}
\renewcommand{\paperspacing}{\doublespacing}

\paperspacing

I propose a quantile-based nonadditive fixed effects panel model to study heterogeneous causal effects.
This model connects to both the standard fixed effects model and the structural quantile regression model. 
It uncovers the heterogeneous causal effects as functions of the unobserved rank variable. The rank is assumed stable over time, which is often more economically plausible than the panel quantile studies that assume individual rank is iid over time.
I provide identification and estimation results, establishing uniform asymptotics of the heterogeneous causal effect function estimator. 
Simulations show reasonable finite-sample performance and show my model complements fixed effects quantile regression.
Finally, I illustrate the proposed methods by examining the causal effect of a country's oil wealth on its military defense spending.



\end{abstract}

\newcommand{\paperspacing}{\onehalfspacing}
\renewcommand{\paperspacing}{\doublespacing}
\paperspacing


\newpage

\section{Introduction}
\label{sec:intro}

Quantile models can be interpreted as random coefficient models in a cross-sectional setting \citep[Section 2.6]{Koenker2005}.
Consider the standard structural quantile model \begin{equation}\label{eqn:structural-simple}
Y_i = \vecf{X}_i' \vecf{\beta}(U_i)
\end{equation}
with random variable $U_i \sim \UnifDist(0,1)$ and the monotonicity condition that $\vecf{x}' \vecf{\beta}(u)$ is increasing in $u$ \textit{for all possible non-random $\vecf{x}$ values}. 
Given the exogeneity assumption $U \independent \vecf{X}$,
this random coefficient model leads to linear conditional quantile function
\begin{equation}\label{eqn:qr}
    \Q_\tau \left( Y \mid \vecf{X}=\vecf{x} \right) = \vecf{x}^\prime \vecf{\beta}_{\mathrm{CQF} }(\tau),
\end{equation}
which can be consistently estimated by \citet{KoenkerBassett1978}.
When there is endogeneity, meaning the 
$\vecf{X}$ and $U$ are statistically dependent, 
the \textit{conditional quantile slope} function $\vecf{\beta}_{\mathrm{CQF} }(\cdot)$ in \cref{eqn:qr} is different from the \textit{structural quantile slope} function $\vecf{\beta}(\cdot)$ in \cref{eqn:structural-simple}.  
\Citet{ChernozhukovHansen2005} use instruments to identify the structural quantile effect in \cref{eqn:structural-simple} in a cross-sectional setting.
With panel data,
I can relax the independence restriction $U \independent \vecf{X}$ to allow unrestricted dependence between $U$ and $\vecf{X}$ without using instruments. 
I extend \cref{eqn:structural-simple} to a panel model and aim to identify and estimate the \textit{heterogeneous causal  effect function} $\vecf{\beta}(\cdot)$.

In this paper, I consider the model
\begin{equation}\label{eqn:model}
        Y_{it}=\vecf{X}_{it}'\vecf{\beta}(U_i)+V_{it},
\end{equation}
where $U_i \sim \UnifDist(0,1)$ is the individual-specific time-invariant unobserved heterogeneity (i.e., ``fixed effect''), 
$V_{it}$ is the idiosyncratic error which will be discussed soon in \cref{sec:ID-pt}, and $U_i$ and the $\vecf{X}_{it}$ are allowed to be arbitrarily dependent. 
Due to the endogeneity of $\vecf{X}_{it}$ and the additive $V_{it}$,
the structural function $\vecf{X}_{it}'\vecf{\beta}(U_i)$ in \cref{eqn:model} does not generally correspond to any conditional quantile function.
Thus, the identification is more complex than showing equivalence to some quantile regression with the right control.
It is important to consider the structural models like \cref{eqn:model}, because the structural function matters for policy and counterfactual analysis, instead of 
the conditional quantile function 
that has only descriptive and predictive meanings without further assumptions.

This model \cref{eqn:model} connects to both the standard fixed effects (FE) model and FE quantile regression (FE-QR) model for the reasons described in detail in \cref{sec:2.1:model:discuss}.
Additionally, similar to \cite{HausmanEtAl2021}, the $V_{it}$ could be considered as measurement error, and the $\vecf{\beta}(\tau)$ has quantile-related interpretation as discussed in \cref{sec:2.1:model:discuss}.

I make four primary contributions.
First, I propose a new quantile-based nonadditive fixed effects model \cref{eqn:model}.
It considers the heterogeneous causal effect as a function of rank variable.
I propose the rank variable to be time-stable, 
for the reasons described later in \cref{sec:2.1:model:discuss}.

Second, 
I identify the heterogeneous causal effects in \cref{eqn:model}.
The structural coefficients measure the causal effects of explanatory variables vector $\vecf{X}$ on the outcome $Y$ for individuals in the population with rank variable value $U=\tau$.
The model deals with endogeneity, meaning the \textit{arbitrary} dependence between regressors and the unobserved heterogeneity (rank variable). 
Identification is achieved by assuming endogeneity is due only to 
fixed effects (not idiosyncratic error).
If the stronger monotonicity assumption of \citet{ChernozhukovHansen2005} is imposed, then the structural parameter of interest fully describes the corresponding quantile structural functions and the related quantile treatment effects.

Third, 
I establish uniform consistency of the coefficient function estimators over all individual rank values 
under a certain rate relation between $n$ and $T$ (the cross-sectional and time series dimensions of the data).
I first estimate the individual-specific coefficients,
relying on established time series OLS consistency results.
Then I implement a sorting approach on a hypothetical outcome at certain $\vecf{x}^{*}$ with the estimated individual coefficients. 
Under a certain rate condition between $n$ and $T$, 
the order of the rank variables can be recovered through the order of the hypothetical outcomes. 
Therefore, 
the estimator of coefficient at any specific rank value is determined
and consistently estimated.

Finally, 
I establish the asymptotic normality of the coefficient function estimator uniformly over all rank values over a proper subset of the unit interval $[0,1]$, 
by applying the functional delta-method to the empirical quantile process.

Although my model is a long panel model, 
a simulation is provided to show the performance of my estimator in the short-$T$ case (for example, $T=3, 4, \ldots, 10$). 
The simulation result is compatible with consistency. 
The simulation also investigates the estimator's performance with various $n/T$ rate relations, which are weaker than the sufficient rate assumed for large-sample theory. 
I also illustrate the performance of my estimator with various sorting point and compare with the FE-QR estimator and standard FE estimator.

\paragraph{Literature}

My model is a \textit{fixed effects} (FE)
approach because it allows arbitrary dependence between $\vecf{X}$ and $U$.
The fact that $U$ follows a standard uniform distribution is a normalization, rather than a restriction.
We can think of $U_i$ as a normalization of the individual fixed effect in the standard FE model \cref{std:FE}, i.e., $U_i=F_\alpha(\alpha_i)$, where $F_\alpha(\cdot)$ is the CDF of $\alpha_i$.
As noted by \citet[p.~286]{Wooldridge2010}, the key difference between the FE approach and correlated random effects (CRE) approach is that the FE approach allows an entirely unspecified relationship between $\alpha_i$ and $\matf{X}_i=(\vecf{X}_{i1},\ldots, \vecf{X}_{iT})$, whereas the CRE approach restricts the dependence between $\alpha_i$ and $\matf{X}_i$ in a substantive way, like restricting the distribution of $\alpha_i$ given $\matf{X}_i$. 
The normalization of $\alpha_i$ alone does not affect the dependence relationship flexibility between $\alpha_i$ and $\matf{X}_i$. 
Therefore, the arbitrary dependence between $\vecf{X}$ and $U$ with normalized $U$ still generally belongs to the FE approach.

My model is similar in spirit to the structural random coefficient interpretation of QR \citep[Section 2.6]{Koenker2005}, instrumental variables quantile regression \citep{ChernozhukovHansen2005,ChernozhukovHansen2006,ChernozhukovHansen2008},
and FE-QR \citep{Canay2011}.
In random coefficient form, my model uses the FE approach, rather than the CRE approach that includes \citet{Chamberlain1982} for panel models and \citet{AbrevayaDahl2008}, \citet{ArellanoBonhomme2016}, and \citet{HardingLamarche2017} for panel quantile models.
\Citet{GalvaoKato2017} provide a survey about using the FE and CRE approaches in panel quantile models.

Like the standard fixed effects model, my model is a \textit{structural model}.
The FE-QR in \citet{Koenker2004} and \citet{KatoEtAl2012} 
are statistical models that summarize conditional distributions of the outcome, 
but require additional assumptions for the coefficient to have a structural interpretation. 
In general, my model does not directly connect with these FE-QR models' conditional quantile function (CQF).
Only when $V_{it}=0$, $u\mapsto \vecf{x}'\vecf{\beta}(u)$ is strictly increasing at all $\vecf{x}$ values, and $\vecf{X}$ is exogenous, i.e., $\vecf{X}$ is independent of $U$, then at time $t$, the cross-sectional $\tau$-CQF of $Y_{it}$ given $\vecf{X}_{it}=\vecf{x}$ is $\vecf{x}'\vecf{\beta}(\tau)$.
My model's object of interest is not the CQF, 
but rather the structural effects.

Almost every FE-QR model (including the initial \citet{Koenker2004} model and later \citet{Canay2011}) requires a ``large-$n$ and large-$T$'' framework.
Alternatively, 
the CRE-QR model can work in a short-$T$ framework at the cost of imposing restrictions on the joint distribution between the observed regressors and the unobserved heterogeneity. 
My model shares the same ``large-$n$ and large-$T$'' flavor as in the FE-QR literature.

Compared to other random coefficient panel data models using the FE approach, 
my model focuses on a different estimand of interest.
\Citet{GrahamPowell2012} and \citet{ArellanoBonhomme2012} contribute a random coefficient panel model using an FE approach in a fixed-$T$ and more general framework that allows multi-dimensional individual fixed effects.
\Citet{GrahamPowell2012} focus on the average partial effect;
\citet{ArellanoBonhomme2012} focus on $\vecf{\beta}_i$ instead of $\vecf{\beta}(U_i)$.
In addition, \citet[Sec.\ 4]{ArellanoBonhomme2012} achieve the identification of $\vecf{\beta}_i$ distribution in a short-$T$ setting by assuming $U_i$ independent of $(V_{i1},\ldots, V_{iT})$ conditional on $\matf{X}_{i}$, using the notation of my \cref{eqn:model}, as well as assuming a particular MA model with independent shock on the error term $V_{it}$.
My model does not make these two assumptions. 
Instead, my model identifies the coefficient $\vecf{\beta}(\cdot)$ as a function of the unobserved rank variable $U$, by making the usual assumptions in quantile models, like monotonicity, continuity, and a scalar rank variable in a large-$T$ setting.

There are some other studies on quantile effects in nonlinear panel data models.
\Citet{ChernozhukovEtAl2013panel} has the advantage of 
considering multiple sources of heterogeneity and works for short-$T$ model, but it is limited to discrete-value regressors. 
My model works for large-$T$ model, but it can generally apply to both continuous-value and discrete-value regressors.
\citeposs{ChernozhukovEtAl2013panel} identification and estimation rely on a ``time homogeneity'' assumption, which is different but non-nested with my model's assumption.
\Citet{Powell2022} studies panel quantile regression with nonadditive fixed effects. 
Both \citeposs{Powell2022} and my paper's rank variables are not iid over time by construction. 
\Citet{Powell2022} can estimate consistently in small-$T$ case, but requires instruments and assumes the same strong monotonicity condition as in \citet{ChernozhukovHansen2005}.
Its estimation and computation of standard errors can be numerically challenging, as pointed out by \citet{BakerQREGPDStata2016}.
My model does not resort to instruments and identification is achieved with a weaker monotonicity assumption.
The estimation and computation of my method is simple and easy.

Unlike difference-in-differences models that focus on a single binary treatment, my model applies to a variety of structural economic models with a variety of regressors, so my model complements the quantile/distributional difference-in-differences literature that includes the work of \citet{AtheyImbens2006}, \citet{CallawayLi2019}, \citet{Ishihara2023}, and \citet{DHaultfoeuilleEtAl2023}, among others.

\Citet{GrahamEtAl2018} contribute a ``fixed effects'' approach in a random coefficient form panel quantile model
in the large-$n$ and small-$T$ framework.
However, the ``fixed effects'' approach of \citet{GrahamEtAl2018} means the dependence between the regressors $\vecf{X}$ and the \textit{random coefficient} $\vecf{\beta}$ is unrestricted, since they assume coefficient as a function of both $\vecf{X}$ and $U$, i.e., $\vecf{\beta}(U, \vecf{X})$; but they still restrict the conditional distribution of unobserved heterogeneity $U$ given regressors $\vecf{X}$, specifically $U\mid \vecf{X} \sim \UnifDist(0,1)$, 
which means for any subpopulation $\vecf{X}=\vecf{x}$, $U \sim \UnifDist(0,1)$, i.e., people do not select $\vecf{X}$ depending on $U$ and there is no endogeneity between $\vecf{X}$ and $U$.
Alternatively, my model's fixed effects approach assumes the dependence between the regressors $\vecf{X}$ and the unobserved heterogeneity $U$ is unrestricted, which is in line with the fixed effects definition; and my model assumes the coefficient as a function of the unobserved heterogeneity $\vecf{\beta}(U)$, which results in the arbitrary dependence between regressors and coefficients.
The two approaches are complementary.

\paragraph{Paper structure and notation} 
\Cref{sec:ID-pt} presents the model and identification results.
\Cref{sec:est} defines the estimator and shows uniform consistency
and the uniform asymptotic normality of the coefficient function estimator. 
\Cref{sec:sim} provides simulation results.
\Cref{sec:emp} provides empirical illustrations.
\Cref{sec:conclusion} concludes.

Acronyms used include those for 
cumulative distribution function (CDF),
conditional quantile function (CQF),
data generating process (DGP),
fixed effects (FE),
ordinary least squares (OLS),
mean squared error (MSE),
quantile regression (QR),
quantile structural function (QSF),
and quantile treatment effect (QTE).
Notationally, 
random and non-random vectors are respectively typeset as, e.g., $\vecf{X}$ and $\vecf{x}$, 
while random and non-random scalars are typeset as $X$ and $x$ (with the exception of certain constants like the time series dimension $T$), 
and random and non-random matrices as $\matf{X}$ and $\matf{x}$; 
that is, upper case denotes a random variable (like $\vecf{X}$ or $U$), whereas lower case denotes a non-random realized value (like $\vecf{x}$ or $u$).
The uniform distribution is written $\UnifDist(a,b)$;
in some cases this stands for a random variable following such a distribution.
Finally, $\weaklyto$ denotes weak convergence.

\section{Model and Identification}
\label{sec:ID-pt}

Consider the structural panel model 
\begin{equation}\label{T>K:idio-out}
    Y_{it}=\vecf{X}_{it}'\vecf{\beta}(U_i)+V_{it}, \quad i=1, \ldots, n, \ t=1, \ldots, T,
\end{equation}
where $Y_{it}$ is the dependent variable for individual $i$ at time $t$,
$\vecf{X}_{it}$ is the corresponding $K$-vector of explanatory variables,
$U_i$ is the individual-specific unobserved heterogeneity normalized to $U_i\sim\UnifDist(0,1)$,
and $V_{it}$ is the idiosyncratic disturbance.
The dependence between $U_i$ and $\vecf{X}_{it}$ is unrestricted.
The coefficient is a function of the unobserved heterogeneity.
For example, 
$\vecf{\beta}(0.5)$ is the structural coefficient vector for an individual with median level unobservable $U_i=0.5$.
The coefficient function $\vecf{\beta}(\cdot) \colon \R \to \R^K $ is deterministic but unknown.
The $\vecf{\beta}(U_i)$ is a random coefficient vector that can vary across individuals, but the source of the randomness is restricted to 
$U_i$.
It captures the heterogeneity in response in unobservables.
In addition, another source of randomness comes from the idiosyncratic error $V_{it}$, which requires large $T$ to learn the individual $\vecf{\beta}(U_i)$ values.

\subsection{Model interpretation}\label{sec:2.1:model:discuss}

This model connects to both the standard FE model and the FE-QR model, and it has a few interpretations.

First, model \cref{T>K:idio-out} assumes a more general functional form than the standard FE model.
Consider the standard FE model
\begin{equation}\label{std:FE}
    Y_{it}= \vecf{X}'_{-1,it} \vecf{\beta}_{-1} +\alpha_i+ V_{it},
\end{equation}
and my model
\begin{equation}\label{eqn:structural:6}
        Y_{it}=\vecf{X}_{-1,it}'\vecf{\beta}_{-1}(U_i)+\beta_0(U_i)+V_{it},
\end{equation}
where $\vecf{X}_{-1,it}$ is the vector of regressors without the constant term, so $\vecf{X}_{it}'=(1,\vecf{X}_{-1,it}')$.
Both the $\alpha_i$ in \cref{std:FE} and $U_i$ in \cref{eqn:structural:6} represent individual fixed effects.
The standard FE model assumes additive fixed effects and homogeneous causal effects $\vecf{\beta}_{-1}$, whereas my model \cref{eqn:structural:6} allows the fixed effects to enter the slope nonseparably and capture the heterogeneous causal effect $\vecf{\beta}_{-1}(U_i)$ (as a function of the unobserved heterogeneity).
When restricting $\vecf{\beta}(U_i)'=(\beta_0(U_i),\beta_1,\beta_2,\ldots)\equiv(\alpha_i,\vecf{\beta}'_{-1})$, model \cref{eqn:structural:6} becomes the standard FE model \cref{std:FE}.

Second, model \cref{T>K:idio-out} complements FE-QR models; for example, see \citet{Canay2011} and references therein.
Consider an additive idiosyncratic error inside the coefficient function, 
\begin{equation}\label{model:generalU+V}
    Y_{it}  =\vecf{X}_{it}'\vecf{\beta}( \overbrace{U_i+ \tilde{V}_{it}}^{\eqqcolon U_{it}}),
\end{equation}
where 
$\tilde{V}_{it}$ is the idiosyncratic error.
It can be more similar to FE-QR model or my model depending on the relative variation / importance of $U_i$ and $\tilde{V}_{it}$.
More specifically, model \cref{model:generalU+V} is close to the FE-QR model when $\tilde{V}_{it}$ dominates $U_i+\tilde{V}_{it}$. 
In an extreme case that $U_i=0$, $U_{it}=0+\tilde{V}_{it}$ is iid over $t$.
Model \cref{model:generalU+V} is close to my model when $U_i$ dominates $U_i+\tilde{V}_{it}$, so $U_{it}$ is stable over time for each individual $i$.

More specifically, 
in model \cref{model:generalU+V}, 
$U_i$ is the individual time-invariant component of unobserved heterogeneity, which can arbitrarily correlate with regressors $\matf{X}_i\equiv(\vecf{X}_{i1},\ldots, \vecf{X}_{iT})$; and $\tilde{V}_{it}$ is the idiosyncratic shock,
which is assumed conditional mean zero given $\vecf{X}_{it}$ and $U_i$, i.e., $\E[\tilde{V}_{it} \mid \vecf{X}_{it}, U_i]=0$.
Consider $U_{it}\coloneqq U_i+\tilde{V}_{it}$ follows a standard uniform distribution on $(0,1)$, 
and the map $u \mapsto \vecf{x}'\vecf{\beta}(u)$ is strictly increasing.
Then $U_{it}$ is considered as the \textit{rank variable}, which represents the rank (quantile) of counterfactual outcome when everyone in the population takes the same $\vecf{x}$ value.

As noted, an important feature of my model is that the rank variable is stable over time, which is often more economically plausible than the time-iid rank assumption made (often implicitly) in the panel quantile regression literature.
The time-iid rank $U_{it}$ means the $U_{it}\sim \UnifDist(0,1)$ is both iid over $i$ and \textit{iid over} $t$.
The time-stable rank $U_{it}$ means the $U_{it} \sim \UnifDist(0,1)$ is iid only over $i$, but has \textit{dependence over} $t$ for each $i$.
The time-stable rank variable idea has been implemented in some studies, including
\citeposs{AtheyImbens2006} quantile difference-in-differences setting and \citeposs{Powell2022} extension of instrumental variables quantile regression to panel data with non-additive FE.
However, to the best of my knowledge, most papers in the panel quantile studies involving a random coefficient form $\vecf{X}_{it}'\vecf{\beta}(U_{it})$ assume a time-iid rank variable, such as Assumption 3.1(a) of \citet{Canay2011} and Assumption 2.1(c) of \citet{ArellanoBonhomme2016}.

The time-stable rank $U_{it}$ is often more economically plausible than the time-iid rank $U_{it}$ in the panel quantile literature.
For example, in the return to education example, individual's innate ability is considered as the unobserved rank variable that determines the rank of outcome earnings. If the individual rank is iid over time, it means that a person whose innate ability is at the 90th percentile in the population (high ability person) in this time period could be arbitrarily at the 10th percentile (the same person becomes very low ability) in the next period, which is not empirically realistic.

Third, similar to \citet{HausmanEtAl2021}, the additive error $V_{it}$ in \cref{T>K:idio-out} can be considered as the measurement error of the outcome variable\footnote{I thank an anonymous reviewer for pointing out this interpretation.}.
That is, we observe $Y_{it}=Y_{it}^*+V_{it}$, where the latent outcome $Y_{it}^{*}$ is generated by $Y_{it}^{*}=\vecf{X}_{it}'\vecf{\beta}(U_i)$
with the same assumptions about rank variable, monotonicity, and continuity as in the general model \cref{T>K:idio-out}.
In this case, the $\vecf{\beta}(\cdot)$ in \cref{T>K:idio-out} can be interpreted in terms of quantile treatment effects and the quantile structural functions for the latent outcome, which will be discussed 
in detail in \cref{sec:2.3:interpret,sec:2.4:connectCQF}.

\subsection{Formal assumptions and results}

Next, I introduce a set of formal assumptions for model \cref{T>K:idio-out}.

\begin{assumption}[iid sampling]\label{A1:iidSampling}
Let $\vecf{Y}_i \equiv (Y_{i1}, \ldots, Y_{iT})' $ 
and matrix $\matf{X}_i \equiv (\vecf{X}_{i1}, \ldots, \vecf{X}_{iT})'$.
The observables $ (\vecf{Y}_i, \matf{X}_i)$
are assumed to be independent and identical distributed (iid) across $i=1, \ldots, n$.
\end{assumption}

\begin{assumption}[Rank variable]\label{A2:rank}
The unobserved individual rank variable $U_i$ is iid across $i$. It follows a standard uniform distribution, $U_i \iid \UnifDist(0,1)$.
\end{assumption}

\begin{assumption}[Monotonicity]\label{A3:monoto}
There exists known $K$-vector $\vecf{x}^*$ in the support of $\vecf{X}_{it}$ 
such that the map $u \mapsto \vecf{x}^{*\prime}\vecf{\beta}(u)$ is strictly increasing on $[0,1]$.
\end{assumption}

\begin{assumption}[Continuity]\label{A4:cts:weakest}
The function $u \mapsto$  $ \vecf{x}^{*\prime} \vecf{\beta}(u)$ is continuous in $u$ on $(0,1)$.
\end{assumption}

\Cref{A3:monoto} and the uniform distribution in \cref{A2:rank} are less restrictive than they seem.
First, consider a generic non-uniform $W_i$ that generates coefficient vector $\vecf{\gamma}(W_i)$.
If its CDF $F_W(\cdot)$ is continuous, and writing $Q_W(\cdot)$ for its quantile function, then by the probability integral transform $\tilde{U}_i=F_W(W_i)\sim\UnifDist(0,1)$, and we can write $\tilde{\vecf{\beta}}(u)=\vecf{\gamma}(Q_W(u))$ so that $Y_{it}=\vecf{X}_{it}'\vecf{\gamma}(W_i) = \vecf{X}_{it}'\tilde{\vecf{\beta}}(\tilde{U}_i)$.
Second, let $Y_i^*\equiv\vecf{x}^{*\prime}\tilde{\vecf{\beta}}(\tilde{U}_i)$ with continuous CDF $F_{Y^*}(\cdot)$.
Again by the probability integral transform, $U_i\equiv F_{Y^*}(Y_i^*) \sim \UnifDist(0,1)$.
By construction, $Y_i^*$ is increasing in $U_i$, and this $U_i$ has the same $\UnifDist(0,1)$ distribution as the $\tilde{U}_i$ above.
Define $m(\cdot)$ such that $U_i=m(\tilde{U}_i)$.
Then, as long as $m(\cdot)$ is invertible, we can define $\vecf{\beta}(u) \equiv \tilde{\vecf{\beta}}(m^{-1}(u))$ and rewrite $Y_{it}=\vecf{X}_{it}'\tilde{\vecf{\beta}}(\tilde{U}_i) = \vecf{X}_{it}'\vecf{\beta}(U_i)$.

Most importantly, the relationship between $U$ and the possibly endogenous regressors $\vecf{X}$ is unrestricted.

\Cref{A3:monoto} is much weaker than the full monotonicity assumption at all possible $\vecf{x}$ values in the instrumental variables quantile regression model of \citet{ChernozhukovHansen2005} (hereafter \citetalias{ChernozhukovHansen2005}).
My model only considers monotonicity at a certain $\vecf{x}^{*}$ value, which suffices for identification, and my model does not impose rank invariance across other $\vecf{x}$ values, but rather focuses on the rank of potential outcomes with everyone counterfactually taking the certain $\vecf{x}^*$ value.
That said, as described later, the interpretation of my $\vecf{\beta}(\cdot)$ is much stronger when monotonicity holds across many or all $\vecf{x}$ values.

The $\vecf{x}^{*}$ is picked for our research interest.
In practice, for example, if $\vecf{X}=(D, \vecf{Z})$ contains a binary treatment variable $D$ and some covariates $\vecf{Z}$, we can let $\vecf{x}^{*}=(1,\vecf{z}^*)$ (or $(0,\vecf{z}^{*})$) if we are interested in the treated (or untreated) counterfactual outcome with everyone taking the covariates value $\vecf{z}^{*}$.%
\footnote{
 I consider the counterfactual distribution in a similar spirit as \citet{ChernozhukovHansen2005} and \citet{AbadieEtAl2002}.
 The difference is that their counterfactual distribution is conditional on covariates $\vecf{Z}=\vecf{z}$, and considering everyone takes the same treatment status $D=d$ value, whereas I consider the counterfactual distribution that both everyone takes the same treatment status $d^{*}$ and takes the same covariates $\vecf{z}^{*}$ value.
 The counterfactual distribution is not conditional on covariates $\vecf{z}$, but unconditionally and hypothetically that everyone takes $\vecf{z}^{*}$ value.
}
If the variable of interest is continuous,
then we can let $\vecf{x}^*$ be the sample average of $\vecf{X}$,
when we are interested in the outcome distribution in a parallel world in which everyone takes the mean value of $\vecf{X}$.
Simulation results indicate the choice of $\vecf{x}^{*}$ does not matter for estimation as long as the population outcome orderings (over individuals) are identical across various $\vecf{x}^*$.

\Cref{A4:cts:weakest} guarantees a unique solution for identification.

\begin{assumption}[Individual slope identification]
\label{A6:idioV}
For each fixed individual $i$, consider its time-series dimension.
That is, suppress the subscript $i$ and write the model as 
\begin{equation}\label{eqn:A6:tsols}
    Y_{t}=\vecf{X}_{t}'\vecf{\beta}+V_{t},\ t=1, \ldots, T,
\end{equation}
where $\vecf{\beta}=\vecf{\beta}_i\equiv\vecf{\beta}(U_i)$ is a fixed coefficient.
Assume (i) no perfect multicollinearity among the $K$ regressors; (ii) $V_t$ is a linear projection error, i.e., for each $t$, $\E(\vecf{X}_tV_t)=\vecf{0}$; 
(iii) $\vecf{\beta}$ is identified. 
\end{assumption}

\Cref{A6:idioV} says the individual coefficient $\vecf{\beta}_i$ is identified as time-dimensional linear projection coefficient, i.e., $ \vecf{\beta}_i=[\E(\vecf{X}_{t}\vecf{X}_{t}')]^{-1}\E \left( \vecf{X}_{t} Y_{t}  \right)$.
There are various lower-level sufficient conditions for part (iii) like covariance stationarity and bounded second moments. 
It can also be denoted as $ \vecf{\beta}_i=[\E(\vecf{X}_{it}\vecf{X}_{it}'\mid  i)]^{-1}\E \left( \vecf{X}_{it} Y_{it} \mid  i  \right)$, where the conditional on individual $i$ means treating individual $i$ as fixed.

Note that conditional on individual $i$ implies conditional on $U_i$, but the other way is not true.
With $\E[\vecf{X}_{it}V_{it}\mid U_i]=\vecf{0}$,
we can also derive from the model \cref{T>K:idio-out} to get $\vecf{\beta}_i=[\E(\vecf{X}_{it}\vecf{X}_{it}'\mid U_i)]^{-1}\E(\vecf{X}_{it}Y_{it}\mid U_i)$.
However, since $U_i$ is unobserved, we cannot connect this to the actual estimation method. 
The estimation in \cref{sec:consist} is conditional on $i$, i.e., treating each individual as fixed and considering its time-dimensional OLS regression.

Note that any additive time-invariant component in $V_{it}$ is absorbed into the intercept term $\vecf{\beta}_0(U_i)$, so $V_{it}$ does not contain any additive time-invariant component.
For example,
if $V_{it}=V_{1i}+V_{2t}$, then the model \cref{T>K:idio-out} becomes 
$ Y_{it}=\underbrace{\beta_0(U_i)+V_{1i}}_{\eqqcolon\gamma_0(U_i) }+\vecf{X}_{-1, it}'\vecf{\beta}_{-1}(U_i)+V_{2t} $ 
with $V_{2t}$ assumed to satisfy \cref{A6:idioV}.

\Cref{A6:idioV} guarantees each individual $\vecf{\beta}_i$ is identified. 
The identification of individual-specific slopes $\vecf{\beta}_1, \ldots, \vecf{\beta}_n$ (from 
\cref{A6:idioV} alone) does not necessarily result in the identification of coefficient functions $\vecf{\beta}(\cdot)$ without further assumptions.

We are interested in the identification and estimation of the structural coefficient function $\vecf{\beta}(\cdot)$ in \cref{T>K:idio-out}.
The model allows for multiple covariates so that
in general the coefficient functions $\vecf{\beta}(\cdot)$ are not necessarily monotone functions.
More specifically, we can identify and estimate the individual's slope $\vecf{\beta}_i \equiv \vecf{\beta}(U_i)$,
but we do not know whether the unobserved rank $U_i$ associated with the observed $\vecf{\beta}_i$ is high or low.
Nor do we know the coefficient function $\vecf{\beta}(\cdot)$ directly.
I use a sorting method in the estimation to recover how the unobserved rank is associated with the observed individual slope,
and therefore, reveal the coefficient function $\vecf{\beta}(\cdot)$.
Only in the degenerate case when $X$ is a scalar,
the coefficient function $\beta(\cdot)$ has to be monotone by \cref{A3:monoto}.

Alternatively, the model can be explained by the following thought experiment. 
Consider there are many counterfactual parallel worlds, 
forcing everyone to take the $\vecf{x}$ values in the support of $\vecf{X}$.
\citetalias{ChernozhukovHansen2005}'s rank invariance is to assume the outcome rank is all the same across all parallel worlds.
My model (when simplified to have no idiosyncratic error) 
assumes the outcome rank in one parallel world ($\vecf{x}^*$) is invariant over time.
It does not restrict the outcome rank in other parallel worlds.

The idiosyncratic error $V_{it}$ is similar in spirit to the notion of rank similarity of \citetalias{ChernozhukovHansen2005}.
Essentially, they allow some ``slippages'' in an individual's potential outcome rank as long as they are exogenous.
Here, the actual outcome ranking can be time-varying, even at the special $\vecf{x}^*$ where monotonicity holds, as long as such variations in $V_{it}$ are exogenous.

\Cref{A1:iidSampling,A3:monoto,A2:rank,A4:cts:weakest,A6:idioV} 
establish the identification of the coefficients at any rank $\tau$.

\begin{theorem}[Identification]\label{thm2:idio:identification}
Under \Cref{A1:iidSampling,A3:monoto,A2:rank,A4:cts:weakest,A6:idioV} and the structural model in \cref{T>K:idio-out},
the coefficients are identified at any rank value.
That is,
for any $\tau \in (0,1)$,
there exists $\vecf{\beta}(\tau)$ in the support of $\vecf{\beta}(U)$ uniquely satisfying
\begin{equation}\label{eqn:thm2}
    \Pr\bigl(Y_{\vecf{x}^*}  \le \vecf{x}^{*\prime} \vecf{\beta}(\tau)  \bigr)=\tau ,
\end{equation}
where $Y_{\vecf{x}^*}\equiv \vecf{x}^{*\prime} \vecf{\beta}(U)$ is defined as the counterfactual outcome in a parallel world with no noise term $V=0$ and everyone takes the $\vecf{x}^{*}$ value.
\end{theorem}

Note that the \cref{thm2:idio:identification} results are not conditional probabilities, but unconditional probability considering the potential outcome in a counterfactual parallel world where everyone in the population takes the $\vecf{x}^{*}$ value.

\subsection{Parameter interpretation}\label{sec:2.3:interpret}

The parameter $\vecf{\beta}(\tau)$ has two immediate interpretations.
First, from the model functional form \cref{T>K:idio-out},
$\vecf{\beta}(\tau)$ can be interpreted as the ceteris paribus effect of $\vecf{X}$ on $Y$ for the individual whose rank is $U_i=\tau$.
More specifically,
for the individual whose rank variable $U_i=\tau$,
his structural function is $Y_t=\vecf{X}_t'\vecf{\beta}(\tau)+V_t$.
If holding $V_t=v$ fixed, and increasing his $\vecf{X}_t$ from $\vecf{X}_t=\vecf{x}_1$ to $\vecf{X}_t=\vecf{x}_2$, this individual's outcome $Y_t$ changes by $(\vecf{x}_2-\vecf{x}_1)'\vecf{\beta}(\tau)$.
Additionally, if $V_t$ contains some function of $\vecf{X}_t$ so that it is impossible to hold fixed $V_t$ while increasing $\vecf{X}_t$, the assumption \cref{A6:idioV}(ii) guarantees we can interpret $\vecf{\beta}(\tau)$ as the marginal effect of $\vecf{X}$ on the mean of $Y$ for the individual with $U_i=\tau$.

Second, from the identification result in \cref{eqn:thm2}, $\vecf{\beta}(\tau)$ is interpreted as the effect of $\vecf{X}$ on the $\tau$-quantile of the counterfactual outcome $Y_{x^{*}}$, provided the monotonicity condition in a neighborhood of $\vecf{x}^{*}$.
That is, $\vecf{\beta}(\tau)$ measures the effect of $\vecf{X}$ on the $\tau$-quantile structural function (QSF) $q(\vecf{x},\tau)$ at $\vecf{x}=\vecf{x}^{*}$.
The QSF is defined by \citet[\S3.1]{ImbensNewey2009} as the $\tau$-quantile of counterfactual outcomes when fixing the observed covariate vector at some value $\vecf{x}$ while letting the unobserved variables vary over their unconditional population distribution.
This is effectively the same as the ``structural quantile function'' in (2.4) of \citet{ChernozhukovHansen2008} or the ``quantile treatment response function'' in (2.2) of \citet{ChernozhukovHansen2005}, who phrase it more explicitly in terms of potential outcomes.
All three papers note that endogeneity prevents estimation of the QSF by conventional quantile regression because then the QSF does not equal the conditional quantile function.
In addition, if we counterfactually set everyone's $V_{it}=0$ and $\vecf{X}_{it}=\vecf{x}_1$, and then change everyone to $\vecf{X}_{it}=\vecf{x}_2$,
the population (unconditional) $\tau$-quantile of the counterfactual outcome changes by $(\vecf{x}_2-\vecf{x}_1)'\vecf{\beta}(\tau)$, which measures the $\tau$-quantile treatment effect (QTE),
i.e., $\tau\textrm{-QTE} \equiv q(\vecf{x}_2,\tau)-q(\vecf{x}_1,\tau) \equiv Q_\tau(Y_{\vecf{x}_2})-Q_\tau(Y_{\vecf{x}_1})=(\vecf{x}_2-\vecf{x}_1)'\vecf{\beta}(\tau)$, provided the monotonicity of $\vecf{x}'\vecf{\beta}(u)$ in $u$ holds at both $\vecf{x}_1$ and $\vecf{x}_2$.
Without full rank invariance at every possible $\vecf{x}$ values, 
this interpretation is limited to any $(\vecf{x}_1, \vecf{x}_2)$ that satisfies the monotonicity condition.

These two interpretations reconcile because 
the $\tau$-quantile of the counterfactual outcome
belongs to the individuals whose $U_i=\tau$ for both $\vecf{X}_{it}=\vecf{x}_1$ and $\vecf{X}_{it}=\vecf{x}_2$.   
So the effect on the $\tau$-quantile of counterfactual outcome (i.e., the second interpretation of $\vecf{\beta}(\tau)$) is identical to the effect on outcome for the $U_i=\tau$ individual (i.e., the first interpretation of $\vecf{\beta}(\tau)$).

Moreover, inspired by \citet{HausmanEtAl2021}, we can interpret $V_{it}$ as the measurement error and $Y_{it}^{*}=\vecf{X}_{it}'\vecf{\beta}(U_i)$ as the true latent outcome; 
then the $\vecf{\beta}(\cdot)$ in my model \cref{T>K:idio-out} has the same interpretation as in a simplified model  $Y_{it}=\vecf{X}_{it}'\vecf{\beta}(U_i)$.
More specifically, $\vecf{\beta}(\cdot)$ can be interpreted as the marginal effect of $\vecf{X}$ on the true latent $Y^{*}$ (and mean of $Y$) 
for the individual whose rank variable value is $U=\tau$.

\subsection{Connection to descriptive CQF model}\label{sec:2.4:connectCQF}

Note that my model \cref{T>K:idio-out} is a \textit{structural} model.
The parameters of interest are the structural effects. 
In other words, if \cref{A3:monoto} is strengthened to hold in a small neighborhood of $\vecf{x}^{*}$ and $V=0$, then $\vecf{\beta}(\tau)$ is interpreted as the effect of $\vecf{X}$ on the $\tau$-quantile structural function $q_Y(\vecf{x}^{*}, \tau)$. 
The QSF is different from the CQF in general, as noted by \citet[page 1488]{ImbensNewey2009}.%
\footnote{
\citet[page 1488]{ImbensNewey2009} define the quantile structural function and write, ``Note that because of the endogeneity of $X$, this [QSF] is in general not equal to the conditional quantile of $g(X,\epsilon)$ conditional on $X=x$, $q_{Y\mid X}(\tau \mid x)$.''
}
Let $V=0$. If \cref{A3:monoto} is strengthened to hold at all $\vecf{x}$ in the support of $\vecf{X}$ as in \citetalias{ChernozhukovHansen2005}, then $\vecf{x}'\vecf{\beta}(\tau)$ is the $\tau$-QSF 
$q_Y(\vecf{x},\tau)$, 
the same as identified in \citetalias{ChernozhukovHansen2005} using instrumental variables.
The differences of the QSF over the values of endogenous regressor yields the quantile treatment effects, i.e., $\tau\textrm{-QTE}=q(\vecf{x}_2,\tau)-q(\vecf{x}_1,\tau)$, provided the QSF is identified at both $\vecf{x}_1$ and $\vecf{x}_2$.

My structural model coincides with the CQF only in a very special case.
If $\vecf{X}$ is exogenous (i.e., $\vecf{X}$ and $U$ are independent), $V_{it}=0$, and monotonicity holds at all $\vecf{x}$, then the $\tau$-quantile of the outcome conditional on $\vecf{X}_{it}=\vecf{x}$ is 
$Q_\tau(Y_{it} \mid \vecf{X}_{it} = \vecf{x})
= \underbrace{Q_\tau(\vecf{X}_{it}'\vecf{\beta}(U)+\overbrace{V_{it}}^{=0} \mid \vecf{X}_{it} = \vecf{x})}_{\vecf{X} \independent U }
= \underbrace{Q_{\tau}(\vecf{x}'\vecf{\beta}(U))}_{\textrm{by \cref{A2:rank} and monotonicity} }
= \vecf{x}'\vecf{\beta}(\tau)$ from the structural model.

\section{Estimation}
\label{sec:est}

First, I define the estimated coefficient function at any rank value. 
Next, I show the uniform consistency and asymptotic normality of the estimated coefficient function as both $T$ and $n$ go to infinity under a certain rate condition.

\subsection{Estimation}
\label{est:largeT}

Define the permutation $\sigma(\cdot)$
to associate the individual ``$i$'' with its order ``$k$'':%
\footnote{
$U_{n:k}$ represents the $k$th order statistic in a sample of size $n$, i.e., 
the $k$th-smallest value among $\{ U_1, \ldots, U_n\}$.
}
\begin{equation}\label{permutation}
    \sigma \left( k \right) \equiv \left\{i: U_{n:k}=U_i \right\} .
\end{equation}
So we can write the coefficient
$ \vecf{\beta} \left( U_{n:k} \right)
= \vecf{\beta} \left( U_{\sigma(k) } \right)$.
Under \cref{A3:monoto}, 
the permutation $\sigma(\cdot)$ can be learned from the $Y^*_i$ ordering, i.e., 
\begin{equation}\label{permutation:Y}
        \sigma \left( k \right) = \left\{i: Y_{n:k}^{*}=Y_i^{*} \right\} .
\end{equation}
because $Y^*_{\sigma(k)}=\vecf{x}^{*\prime}\vecf{\beta}(U_{\sigma(k)})=\vecf{x}^{*\prime}\vecf{\beta}(U_{n:k})=Y^*_{n:k}$.

Then 
$\vecf{\beta} \left( U_{n:\lceil \tau n\rceil} \right)$   is the coefficient associated with the $Y^*_{\sigma(\lceil \tau n\rceil)}=Y_{n: \lceil \tau n\rceil}^{*}$, and $Y_{n: \lceil \tau n\rceil}^{*}$ is the $ \lceil \tau n\rceil$-th smallest value among $\{Y_i^{*}\}_{i=1}^n$,
where 
$U_{n:\lceil \tau n\rceil}$ is the $\lceil \tau n\rceil$th order statistic,
and $\lceil\tau n \rceil $ denotes the ceiling function of $\tau n$, 
i.e., the least integer greater than or equal to $\tau n$.
Because $0<\tau<1$, $ 1 \le \lceil \tau n \rceil   \le n $.

Define the estimated permutation $\hat{\sigma}(\cdot)$
to associate the individual $i$ with its fitted outcome value order $k$:
\begin{equation}\label{permutation:hat}
    \hat{\sigma} \left( k \right) \equiv \left\{i: \hat{Y}_{n:k}^{*}=\hat{Y}_i \right\} ,
\end{equation}
where the fitted outcome is defined just below in \cref{eqn:fitoutcome}.
The estimated permutation $\hat{\sigma} \left( k \right)$ depends on the data and the $\vecf{x}^{*}$ value. 
Later \cref{prop4:betanorm-nrate} shows the estimated order equals the true order with probability approaching one
under a certain rate relation between $n$ and $T$, so $\Pr(\hat\sigma(\cdot)=\sigma(\cdot))\to1$.

Specifically, the estimation of $\vecf{\beta}(\tau)$ consists of four steps.
\begin{steps}
    \item\label{step:general-est}
For each individual,
use its $T$ time periods observations to estimate its individual coefficient:
$ \hat{\vecf{\beta}}_i\equiv \hat{\vecf{\beta}}(U_i) = \left(\matf{X}_i'\matf{X}_i \right)^{-1} \left(\matf{X}_i'\vecf{Y}_i \right) $, $i=1,\ldots,n$.

    \item
Given the estimated individual coefficients 
$\hat{\vecf{\beta}}_1\equiv\hat{\vecf{\beta}}(U_1), \hat{\vecf{\beta}}_2\equiv\hat{\vecf{\beta}}(U_2), \ldots,  \hat{\vecf{\beta}}_n\equiv\hat{\vecf{\beta}}(U_n)  $ from \cref{step:general-est}, 
compute the $n$ fitted outcome values $\hat{Y}^*$ at $\vecf{X}=\vecf{x}^*$,
\begin{equation}\label{eqn:fitoutcome}
    \hat{Y}_i^{*}=\vecf{x}^{*\prime} \hat{\vecf{\beta}}(U_i),\  i=1,\ldots, n.
\end{equation}

    \item\label{step:general-perm}
Sort the fitted outcome values $\{ \hat{Y}_i^* \}_{i=1}^n$ in increasing order:
$\hat{Y}^*_{n:1} \le \hat{Y}^*_{n:2} \le \cdots \le \hat{Y}^*_{n:n}$.
The order of $\hat{Y}^{*}$ is defined as the estimated permutation $\hat{\sigma}(\cdot)$, i.e., $\hat{Y}^*_{n:k}=\hat{Y}^*_{\hat\sigma(k)}$ for $k=1,\ldots,n$.

    \item
Using the $\hat{\vecf{\beta}}_i\equiv\hat{\vecf{\beta}}(U_i)$ from \cref{step:general-est} and the estimated permutation $\hat\sigma(\cdot)$ from \cref{step:general-perm}, the estimator at rank $\tau$ is defined as
\begin{equation}\label{def:beta_hat_generalT}
   \hat{ \vecf{\beta}}(\tau)=  
\hat{\vecf{\beta}}\left( U_{\hat\sigma(\lceil n\tau \rceil)} \right) .
\end{equation}
\end{steps}
I define the estimator in \cref{def:beta_hat_generalT} because 
 $\Pr(\hat\sigma(\cdot)=\sigma(\cdot))\to1$ from \cref{prop4:betanorm-nrate},
that $\vecf{\beta} \left( U_{n:\lceil \tau n\rceil} \right)$ is the coefficient associated with the $Y^*_{\sigma(\lceil \tau n\rceil)}=Y_{n: \lceil \tau n\rceil}^{*}$,
and that $ U_{n:\lceil \tau n\rceil}\pconv \tau$, as $n\to \infty$.

\subsection{Uniform consistency}\label{sec:consist}

I first introduce additional assumptions and then show the uniform consistency of the estimated coefficient function.

Let $\hat{\vecf{\beta}}_T(U_i)  $ denote the time series OLS (TS-OLS) estimator 
for individual $i$'s coefficient $\vecf{\beta}(U_i)  $, 
\begin{equation}\label{est:TS:OLS}
        \hat{\vecf{\beta}}_T(U_i) 
    = \left( \frac{1}{T} \sum_{t=1}^T \vecf{X}_{it}\vecf{X}_{it}' \right)^{-1}
    \left( \frac{1}{T} \sum_{t=1}^T \vecf{X}_{it}Y_{it} \right) .
\end{equation}
There are many large-sample results (\citet[Ch.\ 8.2]{HamiltonText}) for the 
TS-OLS estimator under various assumptions.
For example, given \Cref{A6:idioV} 
and finite second-order moments of $\vecf{X}$ and $V$, the TS-OLS estimator is pointwise consistent for each individual.
That is, $\hat{\vecf{\beta}}_T(U_i) 
= \vecf{\beta}(U_i) +
  \left( \frac{1}{T} \sum_{t=1}^T \vecf{X}_{it}\vecf{X}_{it}' \right)^{-1} 
  \left( \frac{1}{T} \sum_{t=1}^T \vecf{X}_{it}V_{it} \right)\pconv  \vecf{\beta}(U_i) \textrm{ as }  T \to \infty 
     , 
     \  \forall i=1, \ldots, n
.$

I further assume that 
the TS-OLS estimator of each individual-specific coefficient is uniformly consistent 
as $T\to\infty$ and $n\to \infty$.

\begin{assumption}[TS-OLS consistency]\label{A8:TSOLS:CON}
For any positive $\epsilon>0$, 
\begin{equation}\label{A8_uni_consist}
    \Pr\left( \sup_{1\le i \le n} 
    \normbig{ \hat{\vecf{\beta}}_T(U_i)- \vecf{\beta}(U_i) }
    >\epsilon
    \right) \to 0
    \textrm{ as }  n,T \to \infty .
\end{equation}
\end{assumption}

\Cref{A8:TSOLS:CON} implies the pointwise consistency for each individual coefficient.
The norm in \cref{A8_uni_consist} 
is taken as the supremum norm
for simplicity, but it can be interpreted as any $L_p$ norm due to the equivalence of norms in finite-dimensional space.
\cref{A8:TSOLS:CON} implicitly restricts how fast $T$ must grow with respect to $n$. 
A sufficient $n/T$ rate relation is given in the uniform consistency result of the estimated coefficient function $\hat{\vecf{\beta}}(\cdot)$ in \cref{thm5:largeTNconsis}.


\begin{assumption}[Uniform continuity]\label{A7:beta:cts:uniform}
Each element of the coefficient $ \vecf{\beta}(\cdot)$ is a uniformly continuous function on $(0,1)$.
That is, 
writing
$\vecf{\beta}(\cdot)= \left(\beta_1(\cdot), \ldots, \beta_K(\cdot) \right)'$, for any $k=1, \ldots, K$, 
\begin{equation*}
    \forall \epsilon > 0, \exists \  \delta >0 \textrm{ such\ that } \forall x, y \in (0,1),\; \absbig{x-y} \le \delta \implies \absbig{\beta_k(x)-\beta_k(y)} < \epsilon. 
\end{equation*}
\end{assumption}

\Cref{A7:beta:cts:uniform} is a stronger condition than \cref{A4:cts:weakest}.
With the uniform continuous coefficient function assumed in \cref{A7:beta:cts:uniform}
and the uniform consistency 
of the uniform sample quantile function
from \citet[Thm.\ 3, p.\ 95]{ShorackWellner1986},
I establish the uniform consistency of the coefficient function estimator $\hat{\vecf{\beta}}(\cdot)$ on $(0,1)$.

\begin{assumption}\label{A9:LowerBoundBetaDeriv} (Extended monotonicity condition)
Let $g(u)=  \vecf{x}^{*\prime}  \vecf{\beta}(u)$. 
The derivative function $g'(u)$
has a positive lower bound $L$ on $(0,1)$: $\inf_{0<u<1}g'(u)\ge L>0$.
\end{assumption}

The monotonicity condition \cref{A3:monoto} assumes 
the function $g(u)=  \vecf{x}^{*\prime}  \vecf{\beta}(u)$ is strictly increasing on $(0,1)$.
\Cref{A9:LowerBoundBetaDeriv} strengthens this slightly, so $g(u)$ has a strictly positive lower bound of its derivative.
It prevents the situation that the function $g(u)$ is almost flat on some intervals; for example, $g(u)=\sin(u\pi/2)$ satisfies \cref{A3:monoto} because $g'(u)>0$ for all $0<u<1$ but violates \cref{A9:LowerBoundBetaDeriv} because $\lim_{u\uparrow1}g'(u)=0$. 
This restriction is not a strong assumption 
in the sense that 
it still allows
an individual element function of 
the vector-valued coefficient $\vecf{\beta}(\cdot)$ 
to be non-monotone or almost flat.

\begin{assumption}\label{A10:TS-OLS-finiteMean}
For all $T>T_0$, 
$\E \left( T^{\kappa} \norm{\hat{\vecf{\beta}}_{T}(U_i)-\vecf{\beta}(U_i)} \right)  <\infty$,
where $T_0 \in \mathbb{N}$ is some constant and $\kappa>0$ is some positive constant. 
\end{assumption}

\Cref{A10:TS-OLS-finiteMean} helps establish the stochastic order of $\max_{1 \le i \le n} \norm{\hat{\vecf{\beta}}_{T}(U_i)-\vecf{\beta}(U_i)} $ in terms of $n$, which then translates to $T$ using a specified $n/T$ rate relation.
By Theorem A.7 of \citet{LiRacine2007}, \cref{A10:TS-OLS-finiteMean} implies the individual coefficient TS-OLS convergence rate, i.e., $  \hat{\vecf{\beta}}_T(U_i)- \vecf{\beta}(U_i)
         = O_p \left( T^{-\kappa} \right) \textrm{ as }  T \to \infty ,
         \  \forall i=1, \ldots, n. $

\Cref{prop4:betanorm-nrate} shows a sufficient $n/T$ rate relation that guarantees
the true order of the $U_i$ can be discovered from the order of $\hat{Y}^*_i$.
That is, \cref{prop4:betanorm-nrate} implies the estimated permutation is asymptotically the same as the true permutation.

\begin{proposition}[Estimated permutation consistency]\label{prop4:betanorm-nrate}
Suppose
\cref{A1:iidSampling,A2:rank,A3:monoto,A4:cts:weakest,A6:idioV,A7:beta:cts:uniform,A8:TSOLS:CON,A9:LowerBoundBetaDeriv,A10:TS-OLS-finiteMean} hold.
Under the rate relation $n=o \left( T^{\kappa/(3+\delta)} \right)$, 
for some small positive $\delta>0$,
\begin{equation}
        \Pr\left( 
        \hat{\sigma}(\cdot)=\sigma(\cdot)
    \right)
    \to 1 \textrm{ as } n,T\to\infty.
\end{equation}
\end{proposition}

The coefficient estimator defined based on the estimated permutation is uniformly consistent over $(0,1)$.

\begin{theorem}[Uniform consistency]\label{thm5:largeTNconsis}
Under \cref{A1:iidSampling,A2:rank,A3:monoto,A4:cts:weakest,A6:idioV,A7:beta:cts:uniform,A8:TSOLS:CON,A9:LowerBoundBetaDeriv,A10:TS-OLS-finiteMean}, 
as $n$ and $T$ go to infinity with the rate relation 
$n=o( T^{\kappa/(3+\delta)} ) $,
the estimated coefficient function uniformly converges to the true coefficient function,
\begin{equation*}
\sup_{\tau \in (0,1)}  
\normbig{\hat{\vecf{\beta}}(\tau)- \vecf{\beta}(\tau) } = o_p(1).
\end{equation*}
\end{theorem}

The $n/T$ rate relation in \cref{prop4:betanorm-nrate} and \cref{thm5:largeTNconsis} is a \textit{sufficient} rate to guarantee uniform consistency. 
The benchmark rates for FE-QR type of model is $n=o(T^{1/2})$.
The simulation section shows the performance of the proposed estimator in an example DGP with various more relaxed $n/T$ rate relations and with small $T$, like $T=3, \ldots, 10$, in some case.

\subsection{Asymptotic normality }
\label{sec:5.1:inf-AsyNormal}

In this subsection, 
I establish the uniform asymptotic normality of the coefficient function estimator
by applying the functional delta-method to the well-known Donsker's theorem,
assuming differentiability of the coefficient function.

Donsker's theorem \citep[e.g.,][Example 2.1.3]{vanderVaartWellner1996} states the asymptotic Gaussianity of the empirical distribution function. 
Letting $F_n$ be the sequence of the empirical distribution function indexed by sample size $n$, assuming iid sampling
from distribution function $F$,
then
\begin{equation}\label{DonskerThm:general}
    \sqrt{n} \left( F_n -F \right) \weaklyto \mathbb{G} 	\circ F, 
     \ \textrm{ in }   \ell^{\infty}(\bar{\mathbb{R}}), 
    \ \textrm{ as } n \to \infty, 
\end{equation}
where $\mathbb{G} $ is the standard Brownian bridge.\footnote{
The standard Brownian bridge process $\mathbb{G} $ on the unit interval $[0,1]$ is mean zero with covariance function $\Cov[ \mathbb{G}(u), \mathbb{G}(v)]=u \wedge v - u v$,
where $u \wedge v$ denotes $\min \{ u,v \}$.
} 
The limiting process 
is mean zero and has covariance function 
$\Cov \left[ \mathbb{G}_F(u), \mathbb{G}_F(v) \right]$ $=$ $F(u) \wedge F(v)- F(u)F(v)$.
The symbol $\ell^{\infty}(T)$ denotes the collection of all bounded functions $f\colon T \mapsto \R$ with the norm 
$\normbig{f}_{\infty}=\sup_{t\in T}\abs{f(t)}$.

\begin{assumption}\label{A12:betafn:forAsyNormal}
Consider the $K$-vector coefficient function 
$\vecf{\beta}\colon (0,1) \mapsto \R^{K} $.
Assume
each of its element function $\beta_k\colon (0,1) \mapsto \R$, $k=1, \ldots, K$, 
is differentiable with uniformly continuous and bounded derivative.
\end{assumption}

\Cref{A12:betafn:forAsyNormal} is a regularity condition
that assumes the differentiability of the coefficient function.

The uniform asymptotic normality of the coefficient function estimator also holds
under the same sufficient $n/T$ rate relation for its uniform consistency. 
The TS-OLS individual coefficient estimates do not exactly equal to the true coefficient parameters,
but the order of the rank variables can still be discovered asymptotically under the rate condition as shown in \cref{prop4:betanorm-nrate}.
Under the same rate condition, the TS-OLS estimation error goes away even after scaling by $\sqrt{n}$. 
Thus, together with applying the fundamental property in \cref{lemma19:asyNormal} to a standard uniform distribution function, 
we get the uniform asymptotic normality of the coefficient function estimator

\begin{theorem}[Uniform asymptotic normality]\label{thm13:largeTNasynormal}
Under \cref{A1:iidSampling,A2:rank,A3:monoto,A4:cts:weakest,A6:idioV,A7:beta:cts:uniform,A8:TSOLS:CON,A9:LowerBoundBetaDeriv,A10:TS-OLS-finiteMean,A12:betafn:forAsyNormal}, 
as $n,T\to\infty$ with rate relation 
$n=o\left( T^{\kappa/(3+\delta)} \right)$, 
the coefficient function estimator converges to a Gaussian limit,
\begin{equation*}
\sqrt{n}
\bigl(
\hat{\beta}_k (\cdot)- \beta_k (\cdot) 
\bigr)
\weaklyto 
\beta'_k \mathbb{G}, \ \textrm{ in } \ell^{\infty}(0,1),
\end{equation*}
where the limit is mean zero and has covariance function
$ \Sigma(u,v)= \Cov[ (\beta'_k \cdot \mathbb{G})(u), (\beta'_k\cdot\mathbb{G})(v)]= \beta'_k \left( u \right) \beta'_k \left(  v \right) \left( u \land v -uv \right)$, $u,v \in (0,1)$.
\end{theorem}

\Cref{cor14:asyNormal} applies
\cref{thm13:largeTNasynormal} at a fixed $\tau$
to get the pointwise asymptotic normality of the coefficient estimator at any specific rank.

\begin{corollary}[Pointwise asymptotic normality of coefficient estimator]\label{cor14:asyNormal}
Under \cref{A1:iidSampling,A2:rank,A3:monoto,A4:cts:weakest,A6:idioV,A7:beta:cts:uniform,A8:TSOLS:CON,A9:LowerBoundBetaDeriv,A10:TS-OLS-finiteMean,A12:betafn:forAsyNormal} 
for any $\tau \in (0,1)$, 
\begin{equation*}
    \sqrt{n} \bigl( \hat{\beta}_k (\tau) - \beta_k (\tau) \bigr)
\dconv
    \beta'_k \left( \tau \right) \mathbb{G} \left( \tau \right),
    \textrm{ as } n \to \infty, 
\end{equation*}
where $\beta'_k \left( \tau \right) \mathbb{G} \left( \tau \right)$ is a mean-zero Gaussian random variable 
with variance $\tau \left( 1-\tau \right) \left(\beta'_k \left( \tau \right) \right)^2$.
\end{corollary}

\section{Simulation}
\label{sec:sim}

In this section, first, I present the estimator's performance at various $n/T$ relations. 
Second, I consider whether the choice of sorting point matters for the estimates.
Third, I compare my estimator with \citeposs{Canay2011} FE-QR estimator.
Results can be replicated with my code in R \citep{R.core}. All codes are available online.\footnote{\url{https://xinliu16.github.io/}}

\subsection{Varying \texorpdfstring{$n/T$}{n/T} rate}\label{sec:4.1:sim:varynT}

This section illustrates the performance of the estimator at different $n/T$ rate relations.
\cref{thm5:largeTNconsis} provides a theoretical sufficient condition of $n/T$ rate relation to guarantee consistency.
The necessary convergence rate can be much smaller than \cref{thm5:largeTNconsis} provides.
This section uses simulation to check whether the proposed estimator works well in practice with relatively small $T$.

The data generating process (DGP) is
\begin{equation}\label{eqn:sim:4.1}
    Y_{it}= \beta_0(U_i)+\beta_1(U_i)X_{it}+V_{it}, 
\end{equation}
where the regressor $X_{it}=X_{ind, it}+4+\rho U_i$ with $X_{ind, it} \iid N(0,1)$ independent of $(U_i, V_{it})$,
and $\rho$ is some non-negative constant that represents the endogeneity level of the DGP;
the $+4$ is to shift $X_{ind, it}$ to the right by 4 times of its standard deviation to make more than 99.99\% percentage of $X_{it}$ takes positive value in the sample, where the monotonicity condition (of $Y$ in $U$) holds;
the intercept function is $\beta_0(u)=u$; the slope function is $\beta_1(u)=u^2$; 
let $U_i \iid \UnifDist [0,1]$, and the idiosyncratic error $V_{it} \iid N(\mu, \sigma_V^2)= N(0,1)$.

\begin{table}[ht]
    \centering\caption{\label{tab4.1:sim:varynT} MSE of slope estimator } 
 \sisetup{round-precision=3,round-mode=places}
    \begin{threeparttable}
    \begin{tabular}{lllcS[table-format=2.3]S[table-format=2.3]S[table-format=2.3]S[table-format=2.3]cS}
    \toprule
    & &  &   & {$T=n^{1/4}$} & {$T=n^{1/2}$} & {$T=n^{3/4}$}          & {$T=n $} &    \\
    \cmidrule{5-8}  
 $\hat{\beta}_1(0.25)$        &&  $n=   100$  &  &   1.7285009  &  0.1224775  &  0.0310178  &   0.0100735    \\
                              &&  $n=   200$  &  &   0.6432987  &  0.0881695  &  0.0165403  &   0.0051952    \\
                              &&  $n=   500$  &  &   0.3192199  &  0.0487362  &  0.0096771  &   0.0021512    \\
                              &&  $n=  1000$  &  &   0.2659903  &  0.0355368  &  0.0049983  &   0.0009768    \\
                              &&  $n=  2000$  &  &   0.2068939  &  0.0228718  &  0.0027868  &                \\
                              &&  $n=  5000$  &  &   0.1587206  &  0.0134195  &             &                \\
                              &&  $n= 10{,}000$  &  &   0.1486972  &  0.0095093  &             &                \\
    \cmidrule{5-8}  
 $\hat{\beta}_1(0.50)$        &&  $n=   100$  &  &   1.2912381  &  0.1483881  &  0.0381265  &   0.0128576    \\
                              &&  $n=   200$  &  &   0.7815695  &  0.0946917  &  0.0224934  &   0.0063482    \\
                              &&  $n=   500$  &  &   0.6974330  &  0.0555966  &  0.0109769  &   0.0026812    \\
                              &&  $n=  1000$  &  &   0.3321524  &  0.0324294  &  0.0057937  &   0.0011879    \\
                              &&  $n=  2000$  &  &   0.2681620  &  0.0233885  &  0.0037533  &                \\
                              &&  $n=  5000$  &  &   0.2004552  &  0.0157079  &             &                \\
                              &&  $n= 10{,}000$  &  &   0.1525115  &  0.0106901  &             &                \\
    \cmidrule{5-8}  
 $\hat{\beta}_1(0.75)$        &&  $n=   100$  &  &   2.3452817  &  0.1620403  &  0.0478551  &   0.0160400    \\
                              &&  $n=   200$  &  &   1.2896303  &  0.1048440  &  0.0237318  &   0.0077367    \\
                              &&  $n=   500$  &  &   0.7564612  &  0.0651930  &  0.0130435  &   0.0028872    \\
                              &&  $n=  1000$  &  &   0.3790829  &  0.0417468  &  0.0079383  &   0.0015370    \\
                              &&  $n=  2000$  &  &   0.3060419  &  0.0247985  &  0.0041313  &                \\
                              &&  $n=  5000$  &  &   0.2269552  &  0.0158395  &             &                \\
                              &&  $n= 10{,}000$  &  &   0.1825319  &  0.0125885  &             &                \\
\bottomrule
            \end{tabular}
            \begin{tablenotes}
            \item $500$ replications. $x_1^{*}=4.5$. $\rho=1$. Shift 4. $\sigma_V=1$.
            \end{tablenotes}
            \end{threeparttable}
            \end{table}

\Cref{tab4.1:sim:varynT} reports the mean squared error (MSE) of the slope estimators at $0.25$-, $0.5$-, and $0.75$-quantiles with different $n/T$ rate relations $T=n$, $T=n^{3/4}$, $T=n^{1/2}$, and $T=n^{1/4}$.
I let $\rho=1$ to denote the endogeneity and pick the sorting point at $(1, x_1^{*})=(1, 4.5)$.
Vertically, the MSE decreases significantly as $n$ increases from $n=100$ to $n=10{,}000$ in all the $n/T$ rate relation columns.\footnote{The blank place is due to out of software memory in R in the large $n$ and large $T$ cases, like $n=T=10{,}000$.}
The higher the rate, the quicker MSE drops towards zero. For example, for the $\tau=0.5$ slope estimator, the MSE drops from 0.013 to 0.001, as $n$ increases from 100 to 1000 at rate $T=n$; MSE drops from 0.038 to 0.004, as $n$ increases from $100$ to $2000$ at rate $T=n^{3/4}$; and MSE drops from 0.148 to 0.011, as $n$ increases from 100 to $10{,}000$ at rate $T=n^{1/2}$.
Horizontally, at each fixed $n$, the MSE also drops as $T$ increases.

These simulation results are compatible with the consistency of the estimator.
The MSE is decreasing towards zero as $n$ increases even with the slow $T=n^{1/4}$ rate,
where the $T$ is relatively small
(no more than $10$)\footnote{The $T$ ranges from 3 to 10 in the $T=n^{1/4}$ column. Even with the largest $n=10{,}000$, $T=n^{1/4}=10{,}000^{1/4}=10$.
} even though theoretically this is not a fixed-$T$ model.
This estimator can be applied to many common microeconometric setting.

\subsection{Various \texorpdfstring{$x^{*}$}{x*}}\label{sec:4.2:sim:varyXstar}

In this section, I investigate if (and how) the choice of sorting point $x^{*}$ affects the estimates.

Consider the same DGP as in \cref{eqn:sim:4.1}.
I consider five various sorting points $x^{*}$ from 2.5 to 6.5 at a increment of 1.
Note that the [2.5, 5.5] includes the 6.7th- to 93.3th- percentile of a $N(4,1)$ random variable, 
and that $\rho U_i \in [0,1]$.
So [2.5, 6.5] includes more than 86.6\% of possible $X$ values.

Note that if the population order of $Y^{*}=\beta_0(U)+\beta_1(U)x_1^{*}$ at a sorting point $x_1^{*}$ is different from the population order of $Y^{*}=\beta_0(U)+\beta_1(U)x_2^{*}$ at another sorting point $x_2^{*}$, then the sample estimates using $x_1^{*}$ and $x_2^{*}$ are expected to be different. 
Thus, I consider specifically the situation that when $Y_{x_1^{*}}$ and $Y_{x_2^{*}}$ have the same \textit{population order}, and see if the sorting points $x_1^{*}$ and $x_2^{*}$ matter for the \textit{sample} estimates $\hat{\beta}_1(\tau)$.

\begin{table}[ht!]
    \centering\caption{\label{tab4.2:sim:varyXstar:n100T100:shift4:Vsd0.1}  MSE and Bias of $\hat{\beta}_1(\tau) $ } 
\sisetup{round-precision=3,round-mode=places}
    \begin{threeparttable}
    \begin{tabular}{llcS[table-format=-1.3]S[table-format=1.3]cS[table-format=-1.3]S[table-format=1.3]r}
    \toprule
    &     &     & \multicolumn{2}{c}{ $\rho=0$} & &  \multicolumn{2}{c}{ $\rho=1$} \\           
    \cmidrule{4-5} \cmidrule{7-8}
         &     &     & {Bias} &  {MSE} &  & {Bias} & {MSE} &  \\      
    \cmidrule{4-8} 
$\hat{\beta}_1(0.25)$ & $x^{*}=   2.5$ & &  -0.000505  &  0.0006327  &  &   -0.000564 &  0.0006357   &    \\
                       & $x^{*}=   3.5$ & &  0.0001204  &  0.0006037  &  &   0.0003772 &  0.0006263   &    \\
                       & $x^{*}=   4.5$ & &  0.0000605  &  0.0005615  &  &   0.0000332 &  0.0005721   &    \\
                       & $x^{*}=   5.5$ & &  0.0001371  &  0.0005521  &  &   0.0001229 &  0.0005517   &    \\
                       & $x^{*}=   6.5$ & &  0.0000320  &  0.0005329  &  &   0.0000120 &  0.0005368   &    \\
\cmidrule{4-8} 
$\hat{\beta}_1(0.50)$ & $x^{*}=   2.5$ & &  -0.005253  &  0.0026104  &  &   -0.005373 &  0.0027590   &    \\
                      & $x^{*}=   3.5$ & &  -0.005372  &  0.0024574  &  &   -0.005211 &  0.0025463   &    \\ 
                      & $x^{*}=   4.5$ & &  -0.006031  &  0.0023985  &  &   -0.005931 &  0.0024073   &    \\ 
                      & $x^{*}=   5.5$ & &  -0.005341  &  0.0024257  &  &   -0.005842 &  0.0024517   &    \\ 
                      & $x^{*}=   6.5$ & &  -0.005417  &  0.0023877  &  &   -0.005249 &  0.0024252   &    \\ 
\cmidrule{4-8} 
$\hat{\beta}_1(0.75)$ & $x^{*}=   2.5$ & &  -0.012079  &  0.0045089  &  &   -0.012205 &  0.0044587   &    \\ 
                      & $x^{*}=   3.5$ & &  -0.011752  &  0.0043362  &  &   -0.012044 &  0.0044081   &    \\
                      & $x^{*}=   4.5$ & &  -0.012271  &  0.0043605  &  &   -0.011881 &  0.0043529   &    \\
                      & $x^{*}=   5.5$ & &  -0.012236  &  0.0043714  &  &   -0.012391 &  0.0043484   &    \\
                      & $x^{*}=   6.5$ & &  -0.012338  &  0.0043396  &  &   -0.012116 &  0.0043646   &    \\
\bottomrule
            \end{tabular}
            \begin{tablenotes}
            \item $500$ replications. $(n, T)=(100, 100)$. Shift 4. $\sigma_V=0.1$.
            \end{tablenotes}
            \end{threeparttable}
            \end{table}

\cref{tab4.2:sim:varyXstar:n100T100:shift4:Vsd0.1} shows the biases are almost the same (up to 0.001 difference) using a variety of sorting points $x^{*}$ at all quantile levels ($\tau=0.25, 0.5, 0.75$).
It illustrates that the sorting point does not matter for sample estimates, as long as the population orders are identical across various sorting points. 
\cref{tab4.2:sim:varyXstar:n1000T1000:shift0} in \Cref{sec:B.2} includes extra simulation results for the DGP without shifting $X$ and pick various sorting points $x^{*}=0, 0.5, 1, 1.5, 2, 2.5$. That is, half of the unshifted $X$ are negative values and the population orders are different at negative $x^{*}$ from that at positive $x^{*}$ values.
The results and conclusion are the same as above that the sample estimates $\hat{\beta}_1(\tau)$ do not rely on the choice of sorting points $x^{*}$, wherever the population orders of $Y^{*}$ are identical across these various $x^{*}$.

\Cref{sec:B.2} includes extra simulation results with larger sample $(n,T)=(1000,1000)$ and specifically small $T$, such as $(n,T)=(1000, 10)$ case.
\cref{tab4.2:sim:varyXstar:n1000T1000:shift4:Vsd0.1} shows with large sample size $(n,T)=(1000,1000)$, my estimator performs very well with up to 0.001 bias and less than 0.0005 MSE at all quantile levels $\tau=0.25, 0.5, 0.75$.
\cref{tab4.2:sim:varyXstar:n1000T10:shift4:Vsd0.1} shows even in the $T=10$ very short-$T$ case, 
my estimator has up to 0.005 bias and the estimates do not vary much (up to 0.009 estimates difference) across different $x^{*}$ values.

\subsection{Compare with FE-QR}\label{sec:4.4:sim:FEQR}

This section compares my estimator and \citeposs{Canay2011} FE-QR estimator in various situations.

Consider the DGP that
\begin{equation}
\begin{split}
    Y_{it}&= \beta_0(U_{it})+\beta_1(U_{it})X_{it}, \\
    U_{it}&\equiv F(U_i+\tilde{V}_{it})
\end{split}
\end{equation}
where the regressor $X_{it}=X_{ind, it}+4+\rho U_i$ with $X_{ind, it} \iid N(0,1)$ independent of $(U_i, \tilde{V}_{it})$,
and $\rho$ is some non-negative constant that represents the endogeneity level of the DGP;
specifically, I consider $\rho=0$ (no endogeneity) and $\rho=1,3,10$ as endogeneity level increases.
The intercept function is $\beta_0(u)=u$; the slope function is $\beta_1(u)=u^2$; 
let $U_i \iid \UnifDist (0,1)$, and the idiosyncratic error $\tilde{V}_{it} \iid N(0, \sigma_V^2)$.
Since the $U_i+\tilde{V}_{it}$ may take value outside $[0,1]$, I normalize it to be on $[0,1]$ to represent the rank variable. 
That is, I consider the CDF of the random variable $U_i+\tilde{V}_{it}$, i.e., let $ F(U_i+\tilde{V}_{it})\sim \UnifDist(0,1)$ and $\beta_0(\cdot)$ and $\beta_1(\cdot)$ are  functions of $F(U_i+\tilde{V}_{it})$.

I take the sorting point at $x^{*}=4$, which is around the center of $X_{it}$, as $X_{ind, it}+4 \sim N(4,1)$ and $\rho U_i \in [0,\rho]$. 
As shown in \cref{sec:4.2:sim:varyXstar}, the choice of sorting points $x^{*}$ does not matter for sample estimates, as long as the population orders of $Y^{*}$ are identical across these $x^{*}$ values.

I consider three cases $\sigma_V=0.01$, $0.1$, and $1$, respectively. 
The third case ($\sigma_V=1$) is towards the Canay FE-QR situation that the time-iid component $\tilde{V}_{it}$ dominates $U_i+\tilde{V}_{it}$. 
The first case ($\sigma_V=0.01$) is towards the situation that rank variable is time-persistent, i.e., $U_i$ dominates the $U_i+\tilde{V}_{it}$.
The second case $\sigma_V=0.1$ is in between and is a time-stable rank variable case.

\begin{table}[htbp]
     \fontsize{11}{12}\selectfont
    \centering\caption{\label{tab4.4:sim:n100:T100}  MSE and Bias of $\hat{\beta}_1(\tau) $ } 
\sisetup{round-precision=3,round-mode=places}
\begin{adjustbox}{width=0.9\columnwidth,center}
    \begin{threeparttable}
    \begin{tabular}{llccS[table-format=-1.3]S[table-format=1.3]cS[table-format=-1.3]S[table-format=1.3]c}
    \toprule
 $\tilde{V}_{it} \sim N(0, \sigma_V^2)$     &               &                        &     & \multicolumn{2}{c}{My model ($x^{*}=4$)   }   & & \multicolumn{2}{c}{FE-QR \citep{Canay2011}   }       &      \\
    \cmidrule{5-6} \cmidrule{8-9} 
    &               &                        &     &    {Bias}    &       {MSE}        & &     {Bias}     &       {MSE}         &  \\      
     
\cmidrule{5-9} 
$\sigma_V=0.01$ & $\rho=  0$   & $\hat{\beta}_1(0.25)$  &     &  0.0001900  &  0.0004816  & &   0.2121819 &  0.0466900   &    \\
&               & $\hat{\beta}_1(0.50)$  &     &  -0.005764  &  0.0023804  & &   -0.000907 &  0.0020398   &    \\
&               & $\hat{\beta}_1(0.75)$  &     &  -0.012423  &  0.0043033  & &   -0.287448 &  0.0843232   &    \\
\cmidrule{5-9} 
& $\rho=  1$    & $\hat{\beta}_1(0.25)$  &     &  0.0001739  &  0.0004814  & &   0.1908570 &  0.0385585   &    \\
&               & $\hat{\beta}_1(0.50)$  &     &  -0.005470  &  0.0023817  & &   0.0061173 &  0.0019000   &    \\
&               & $\hat{\beta}_1(0.75)$  &     &  -0.011856  &  0.0042850  & &   -0.269171 &  0.0737943   &    \\
\cmidrule{5-9} 
& $\rho=  3$    & $\hat{\beta}_1(0.25)$  &     &  0.0002391  &  0.0004967  & &   0.2004619 &  0.0423133   &    \\
&               & $\hat{\beta}_1(0.50)$  &     &  -0.004867  &  0.0024617  & &   0.0387201 &  0.0028684   &    \\
&               & $\hat{\beta}_1(0.75)$  &     &  -0.012447  &  0.0044571  & &   -0.248375 &  0.0627387   &    \\
\cmidrule{5-9} 
& $\rho= 10$    & $\hat{\beta}_1(0.25)$  &     &  0.0002783  &  0.0005105  & &   0.2522033 &  0.0646985   &    \\
&               & $\hat{\beta}_1(0.50)$  &     &  -0.003864  &  0.0029467  & &   0.0721135 &  0.0061763   &    \\
&               & $\hat{\beta}_1(0.75)$  &     &  -0.006820  &  0.0069500  & &   -0.231391 &  0.0544464   &    \\
\cmidrule{2-9} 
$\sigma_V=0.1$ & $\rho=  0$   & $\hat{\beta}_1(0.25)$  &     &  0.0109790  &  0.0013434  & &   0.1816793 &  0.0339376   &    \\
&               & $\hat{\beta}_1(0.50)$  &     &  0.0051668  &  0.0044040  & &   -0.030949 &  0.0022330   &    \\
&               & $\hat{\beta}_1(0.75)$  &     &  -0.009501  &  0.0083931  & &   -0.240430 &  0.0592632   &    \\
\cmidrule{5-9} 
& $\rho=  1$    & $\hat{\beta}_1(0.25)$  &     &  0.0107730  &  0.0016992  & &   0.1108493 &  0.0131660   &    \\
&               & $\hat{\beta}_1(0.50)$  &     &  0.0079887  &  0.0065227  & &   -0.037274 &  0.0025700   &    \\
&               & $\hat{\beta}_1(0.75)$  &     &  -0.002758  &  0.0113306  & &   -0.209409 &  0.0451596   &    \\
\cmidrule{5-9} 
& $\rho=  3$    & $\hat{\beta}_1(0.25)$  &     &  0.0111221  &  0.0020721  & &   0.0356626 &  0.0020265   &    \\
&               & $\hat{\beta}_1(0.50)$  &     &  0.0067142  &  0.0094172  & &   -0.001412 &  0.0012026   &    \\
&               & $\hat{\beta}_1(0.75)$  &     &  0.0076405  &  0.0249752  & &   -0.166035 &  0.0286979   &    \\
\cmidrule{5-9} 
& $\rho= 10$    & $\hat{\beta}_1(0.25)$  &     &  0.0694300  &  0.0336007  & &   0.1177622 &  0.0148625   &    \\
&               & $\hat{\beta}_1(0.50)$  &     &  0.0853712  &  0.0733586  & &   0.0639205 &  0.0052415   &    \\
&               & $\hat{\beta}_1(0.75)$  &     &  -0.026679  &  0.0776072  & &   -0.131956 &  0.0184508   &    \\
\cmidrule{2-9} 
$\sigma_V=1$ & $\rho=  0$   & $\hat{\beta}_1(0.25)$  &     &  0.2031959  &  0.0585703  & &   0.0078521 &  0.0002206   &    \\
&               & $\hat{\beta}_1(0.50)$  &     &  0.0705810  &  0.0259877  & &   -0.003678 &  0.0005692   &    \\
&               & $\hat{\beta}_1(0.75)$  &     &  -0.168066  &  0.0521751  & &   -0.004752 &  0.0008283   &    \\
\cmidrule{5-9} 
& $\rho=  1$    & $\hat{\beta}_1(0.25)$  &     &  0.2223245  &  0.0724398  & &   -0.032061 &  0.0011840   &    \\
&               & $\hat{\beta}_1(0.50)$  &     &  0.1061118  &  0.0440824  & &   0.0244318 &  0.0012359   &    \\
&               & $\hat{\beta}_1(0.75)$  &     &  -0.141973  &  0.0540637  & &   0.0393636 &  0.0024323   &    \\
\cmidrule{5-9} 
& $\rho=  3$    & $\hat{\beta}_1(0.25)$  &     &  0.2869131  &  0.1252145  & &   -0.070942 &  0.0052701   &    \\
&               & $\hat{\beta}_1(0.50)$  &     &  0.1366515  &  0.0566272  & &   0.0641793 &  0.0048235   &    \\
&               & $\hat{\beta}_1(0.75)$  &     &  -0.229447  &  0.0851543  & &   0.0979893 &  0.0104279   &    \\
\cmidrule{5-9} 
& $\rho= 10$    & $\hat{\beta}_1(0.25)$  &     &  0.3806054  &  0.1712474  & &   -0.055123 &  0.0037959   &    \\
&               & $\hat{\beta}_1(0.50)$  &     &  0.0226625  &  0.0231031  & &   0.0735467 &  0.0064529   &    \\
&               & $\hat{\beta}_1(0.75)$  &     &  -0.369314  &  0.1641765  & &   0.0991469 &  0.0108964   &    \\
\bottomrule
            \end{tabular}
            \begin{tablenotes}
            \item $500$ replications. $(n, T)=(100, 100)$. Shift 4.
            \end{tablenotes}
            \end{threeparttable}
            \end{adjustbox}
\end{table}

\Cref{tab4.4:sim:n100:T100}
reports the bias and MSE of my estimator and \citeposs{Canay2011} FE-QR estimator.
In the first case $\sigma_V=0.01$,
my estimator has much smaller bias and smaller MSE than FE-QR estimator at almost all quantile levels ($\tau =0.25, 0.5, 0.75$), especially when the DGP has endogeneity. 
For example, with high endogeneity $\rho=10$,
my estimator has 0.000 bias and 0.001 MSE for $\tau=0.25$ quantile, in contrast to FE-QR estimator has a 0.252 bias and 0.065 MSE.

In the middle case that $\sigma_V=0.1$, my estimator performs comparably to \citeposs{Canay2011} FE-QR estimator.
In the no/low endogeneity case ($\rho=0,1$), my estimator has a smaller bias and MSE than FE-QR estimator.
In the high endogeneity case ($\rho=10$), my estimator usually has a smaller bias, but bigger MSE than FE-QR estimator.
For example, my estimator has 0.069 bias and 0.034 MSE for $\hat{\beta}_1(0.25)$, in contrast to FE-QR estimator has a 0.118 bias and 0.015 MSE.

The FE-QR estimator performs much better with  $\sigma_V=1$ than with smaller $\sigma_V$. 
Only in cases at $\tau=0.5$ with high endogeneity (like $\rho=10$), my estimator has a smaller bias and smaller MSE than FE-QR estimator. 
In most other cases ($\tau=0.5$ with mild endogeneity and $\tau=0.25, 0.75$ with various endogeneity), the FE-QR estimator has a much smaller bias and MSE than my estimator.

\cref{tab4.4:sim:n100:T100} shows my model complements FE-QR. The FE-QR performs poorly once $\sigma_V$ is small enough, but it outperforms my approach once $\sigma_V$ is large enough. 
So in practice we choose between my approach and the FE-QR approach depending on whether we think the $U_i$ or $\tilde{V}_{it}$ is a bigger source of variation.

\cref{sec:B.4} shows the $(n,T)=(1000,10)$ results parallel to \cref{tab4.4:sim:n100:T100}. 
The result patterns are the same.

Moreover, my estimator is computationally simple and fast relative to Canay's FE-QR estimator, 
especially with large $n$ and large $T$.
It takes 10 seconds for my estimator to run 500 replications with four endogeneity cases and a sample size of sample $(n,T)=(250, 250)$, but it takes 4 minutes to run the same for FE-QR estimator.
It is infeasible to compute Canay's FE-QR estimator with 500 replications and sample size $(n,T)=(500,500)$, whereas it only takes 25 seconds for my estimator in the same setting.

\section{Empirical illustration}
\label{sec:emp}

To apply the proposed method to an empirical example, 
I study 
the causal effect of a country's oil wealth on its military defense burden
using the data from \citet{CotetTsui2013}. 
There are many studies debating the effect of oil wealth on political conflict.
On one hand,
resource scarcity triggers conflict, so resource abundance should mitigate regime instability.
On the other hand, 
political instability in the Middle East and some developing countries with rich oil resources makes people arrive at the ``oil-fuels-war'' conclusion.
The political violence in a regime can be measured in multiple ways, for example, the onset of civil war.
Additionally, political conflict may not necessarily lead to regime shift or war.
Military defense burden is also considered as a barrier to political conflict. 
My empirical example studies the effect of oil wealth on military defense spending.

First, I provide point estimates of the effect of oil wealth on military defense spending (as a ratio of GDP)
for countries at various rank levels $\tau$.
The rank variable represents a country's unobserved general propensity of having military spending.
Second, I present bootstrap standard error for the estimated effects.%
\footnote{
Details of bootstrap algorithm for computing standard error are presented in \cref{appx:bs-se-algorithm}.
}

\Citet{CotetTsui2013} use the standard fixed effect method to study the effect of oil wealth on political violence.
They provide several measures of 
oil wealth, such as oil reserves or oil discoveries, 
and several measures related to
political violence, such as the onset of civil war or the military defense spending as a ratio of GDP.
They find that 
after controlling for the country fixed effect, 
oil wealth has no effect on civil conflict onset or military spending.
I investigate if there is effect heterogeneity across the unobserved military spending rank variable, although the effect might be statistically insignificant as \cite{CotetTsui2013} find.

Consider the structural model 
\begin{equation}
    \log( \textit{DEFENSE}_{it} ) 
    = \beta(U_i) \times \log( \textit{OILWEALTH}_{it} ) + \vecf{X}_{it}'\vecf{\gamma}(U_i)
    + V_{it} .
\end{equation}
The outcome variable is measured as (the log of the percentage points of) the ratio of military defense spending to GDP. 
The variable of interest is 
$\log( \textit{OILWEALTH}_{it} )$,
the log of 
dollar-valued inflation-adjusted oil wealth per capita.
The parameter $\beta(U_i)$ is the causal effect of oil wealth on military spending.
The unobserved country rank variable $U_i$
denotes country $i$'s general propensity to have military spending.
The effect 
can differ across countries with different $U_i$.
The covariate vector $\vecf{X}_{it}$ contains country characteristics
including economic growth and population.
$V_{it}$ is the idiosyncratic error, 
which is assumed exogenous so that TS-OLS is consistent.
The goal is to estimate $\beta(\tau)$ for various $\tau$
rank.

The data is from \citet{CotetTsui2013}. 
My sample includes 45 countries over years 1988--2003.
The military defense burden outcome variable is measured as a log of percentage points. 
For example, 
if a country's military spending is 3.2\% of its GDP, 
then its outcome variable value is $\log(3.2)$.
Oil wealth is measured as log oil value per capita.\footnote{
\Citet{CotetTsui2013} instead use the log of oil wealth per capita \emph{divided by $100$}.
My coefficient estimates can be multiplied by $100$ to compare directly with those of \citet{CotetTsui2013}.
} 
For example, 
a country with 6 million dollars per capita oil reserve wealth is counted as
$\log( \numnornd{6000000})=6.8$ as its $\log(\textit{OILWEALTH}_{it})$ value.
The effect $\beta(U_i)$ measures the elasticity given a 1\% increase in oil wealth per capita, i.e., the percent effect on the percentage points of the military spending to GDP ratio.

The two other covariates included in my application are the country's economic growth and population density.
Economic growth is measured as the country's annual GDP growth rate.
For example,
if a country's annual GDP growth is 3\%,
then the value of economic growth is $0.03$.
The population density is measured by the log of the country's total population.\footnote{
\Citet{CotetTsui2013} use log of the country's population divided by 100 as the population density.
}

\begin{table}[htbp]
    \centering\caption{\label{tab:empirical}  Empirical results for effect of oil wealth on military defense spending.}
    \sisetup{round-precision=2,round-mode=places, table-format=-3.2}
    \begin{threeparttable}
        \begin{tabular}[c]{S[table-format=1.2]c
            S[table-format=2.3]S[table-format=2.3]c
            S[table-format=2.3]S[table-format=2.3]c
            S[table-format=2.3]S[table-format=2.3]cr}
            \toprule
            && \multicolumn{2}{c}{My model} && \multicolumn{2}{c}{FE-QR} && \multicolumn{2}{c}{standard FE} \\
            \cmidrule{3-4}     \cmidrule{6-7}  \cmidrule{9-10}
            $\tau$       
            &&  {$\hat{\beta}_\tau$}     &  {$\mathrm{SE}(\hat{\beta}_\tau)$}     
            &&  {$\hat{\beta}_\tau$}     &  {$\mathrm{SE}(\hat{\beta}_\tau)$}  
            &&  {$\hat{\beta}$}           &  {$\mathrm{SE}(\hat{\beta})$} \\
            \midrule
       0.25   &&    -0.339435   &   0.2954894  &&   0.0302177   &   0.1380579  &&        &    \\
        0.5   &&    -0.176758   &   0.2616236  &&   0.0420494   &   0.1045064  &&    0.0436717   &   0.0625423      \\
       0.75   &&    -0.394243   &   0.2299444  &&   0.0488585   &   0.1093188  &&        &    \\
        \bottomrule
        \end{tabular}
        \begin{tablenotes}
            \item  100 bootstrap replications for bootstrap estimator of standard error.
        \end{tablenotes}
    \end{threeparttable}
\end{table}

\Cref{tab:empirical} presents the coefficient estimates at rank values $\tau=\{0.25, 0.5, 0.75\}$
quantile 
using sorting point of the sample mean $\vecf{x}^{*}=1/(nT)\sum_{i=1}^n \sum_{t=1}^T \vecf{X}_{it}$ including both the regressor of interest $\log( \textit{OILWEALTH})$ and other covariates.
The interpretation of the rank variable order is associated with the sorting points. 
The robustness of estimates using nearby sorting points $\vecf{x}^*$ is presented in \cref{appx:emp:x-star}.
For comparison, \cref{tab:empirical} also presents \citeposs{Canay2011} FE-QR estimates and the standard FE estimates.

For all three models, the estimated effects are statistically insignificant.
This is consistent with the literature \citep[e.g.,][]{CotetTsui2013,FearonLaitin2003}.
The existing macro-level studies of the oil-fuels-war hypothesis find that the effect is significant in a pooled OLS setting.
The inclusion of country fixed effects eliminates the statistical association between oil and civil war, and the statistical significance disappears.

Despite the statistical insignificance, I interpret the estimates economically for illustration.
Assuming my model specification, at $\tau=0.5$, the estimated coefficient $-0.18$ means: for a country with median propensity for military spending, a $1\%$ increase in oil wealth per capita is estimated to cause a $0.18\%$ decrease in the percentage points of military spending as a ratio of GDP.
Using Canay's FE-QR model specification, the estimate at $\tau=0.5$ is instead $0.04$, which is close to the ``mean'' effect using the standard FE model.
Of course, different model specifications make very different assumptions.
The standard FE model assumes a homogeneous slope across units, and Canay's FE-QR model allows slope heterogeneity assuming time-iid rank, whereas my model assumes time-invariant rank.
Except in trivial cases, my model and Canay's cannot both be correctly specified, so it is not surprising that the estimates differ economically.

\section{Conclusion}
\label{sec:conclusion}

In this paper,
I propose a structural quantile-based nonadditive fixed effects panel model with 
the goal of discovering heterogeneous causal effects while accounting for endogeneity.
This model generalizes the standard FE model and complements the FE-QR model.
The fixed effects enter the intercept and slopes nonseparably.
The causal effect is heterogeneous in the unobserved rank variable, which represents the rank of the counterfactual outcome at certain $\vecf{x}^{*}$ values that policy-makers are interested in. 
The rank variable is assumed to be time-stable, which often makes more economic sense than the time-iid rank variable in panel quantile regression literature. 
I develop several theoretical results, including identification, uniform consistency, and uniform asymptotic normality of the coefficient estimator. 
Simulation shows the compatibility with consistency even with small-$T$ and at weaker $n/T$ rate relations, that the estimates do not depend on choice of sorting point, and the comparison between my estimator and a popular FE-QR estimator.

\section*{Acknowledgments }
I thank the editor, associate editor, and anonymous reviewers for their helpful comments that greatly improved this paper. 
I'm grateful to David Kaplan for his patient and enthusiastic guidance, encouragements and discussions. I also thank Zack Miller, Shawn Ni, Alyssa Carlson, Rachael Meager, seminar participants at University of Glasgow, University of Amsterdam, Washington State University, University of Connecticut, as well as the participants at 2020 EWMES, 2021 AMES, 2022 MEG, and 2022 SEA for helpful conversations and comments on this project. All errors are mine.

\section*{Disclosure statement}
The author reports there are no competing interests to declare.

\Supplemental{\newpage}{%
\singlespacing
\bibliographystyle{chicago}
\bibliography{_bib}
}

\onehalfspacing

\appendix

\numberwithin{equation}{section}
\numberwithin{theorem}{section}
\numberwithin{assumption}{section}

\clearpage 

\setcounter{page}{1}

\begin{appendices}

This online appendix provides supporting materials for the paper ``A quantile-based nonadditive fixed effects model''.
\cref{sec:app-pfs} provides proofs for the theorems and propositions in Section 2--4.
\cref{sec:appx:addproof} provides additional proofs for lemmas which are used in \cref{sec:app-pfs}.
\cref{sec:appx:sim} provides additional simulation results.

\section{Main Proofs}
\label{sec:app-pfs}

\begin{proof}[Proof of \cref{thm2:idio:identification}]
Under \cref{A6:idioV}, each $\vecf{\beta}_i$ is identified, 
i.e., $ \vecf{\beta}_i=[\E(\vecf{X}_{it}\vecf{X}_{it}'\mid i)]^{-1}\E(\vecf{X}_{it}Y_{it}\mid i)$.
Under \cref{A1:iidSampling,A3:monoto,A2:rank,A4:cts:weakest}, 
\begin{align*}
    \Pr \left( \vecf{x}^{*\prime} \vecf{\beta}(U)  \le \vecf{x}^{*\prime} \vecf{\beta}(\tau)  \right)
    &= \Pr \left( U \le \tau \right) \quad  \textrm{By} \  \cref{A3:monoto} \\
    &=\tau. \quad \textrm{By} \  \cref{A2:rank}
\end{align*}
Therefore, 
$\vecf{\beta}(\tau)$ is a solution to
$ \Pr\left(\vecf{x}^{*\prime} \vecf{\beta}(U)  \le \vecf{x}^{*\prime} \vecf{\beta}(\tau)  \right)=\tau$. 
The uniqueness of $\vecf{\beta}(\tau)$ comes from the strictly increasing of $\vecf{x}^{*\prime} \vecf{\beta}(u)$ in \cref{A3:monoto} and its continuity in \cref{A4:cts:weakest}.
\end{proof}

\begin{proof}[Proof of \cref{prop4:betanorm-nrate}]
I want to show 
\begin{equation}\label{eqn:pw}
        \Pr\left( \frac{ \min\{Y_{n:2}^{*} - Y_{n:1}^{*}, \ldots, Y_{n:n}^{*} - Y_{n:n-1}^{*} \} }{2}
    > \max_{1\le i \le n} \abs{\hat{Y}_{i}^{*} -Y_i^{*} } \right)
    \to 1 \textrm{ as } n,T\to\infty.
\end{equation}

Let $W$ denote the event that 
\begin{equation}
    W=\left\{ \frac{ \min\{Y_{n:2}^{*} - Y_{n:1}^{*}, \ldots, Y_{n:n}^{*} - Y_{n:n-1}^{*} \} }{2}
    > \max_{1\leq i \leq n} \absbig{\hat{Y}_{i}^{*} -Y_i^{*} } \right\} .
\end{equation}

In the event, the left hand side random variable $ \min\{Y_{n:2}^{*} - Y_{n:1}^{*}, \ldots, Y_{n:n}^{*} - Y_{n:n-1}^{*} \}$ only relates with $n$ rate. 
The right hand side random variable $\max_{1\leq i \leq n} \absbig{\hat{Y}_{i}^{*} -Y_i^{*} } $ relates with both $n$ rate and $T$ rate. 
We want to show there exists a sequence of small $\epsilon_n >0 $ such that 
\begin{equation}
    \Pr \left(\frac{\min\{Y_{n:2}^{*} - Y_{n:1}^{*}, \ldots, Y_{n:n}^{*} - Y_{n:n-1}^{*} \} }{ 2 \normbig{\vecf{x}^{*} } } > \epsilon_n
    > \frac{ \absbig{\hat{Y}_{i}^{*} -Y_i^{*} } }{\normbig{\vecf{x}^{*} }}, \ \forall i=1, \ldots, n
    \right) \to 1
\end{equation}
as $n,T \to \infty $ under the $n$ and $T$ rate relation.
This is a sufficient but not necessary condition for \cref{eqn:pw}.

Let 
\begin{equation}
    A= \left\{ 
    \frac{\min\{Y_{n:2}^{*} - Y_{n:1}^{*}, \ldots, Y_{n:n}^{*} - Y_{n:n-1}^{*} \} }{ 2\normbig{\vecf{x}^{*} } } > \epsilon_n \
    \right\}
\end{equation}
and 
\begin{equation}
    B= \left\{ 
     \epsilon_n 
     > \frac{ \absbig{\hat{Y}_{i}^{*} -Y_i^{*} } }{\normbig{\vecf{x}^{*} }}, \ \forall i=1, \ldots, n
    \right\}. 
\end{equation}
I want to show 
\begin{equation*}
    \Pr(A) \to 1
    \textrm{ as } n \to \infty,
\end{equation*}
and for small $\delta>0$,
\begin{equation*}
    \Pr(B) \to 1
    \textrm{ as }  n,T \to \infty \textrm{ with } n=o\left( T^{\kappa/(3+\delta)} \right) .
\end{equation*}
If these both hold, then
\begin{equation}
    \Pr(A \textrm{ and } B) \ge \Pr(A)- \Pr(\bar{B}) \to 1,
    \textrm{ as }  n,T \to \infty, \textrm{ with } n=o\left( T^{\kappa/(3+\delta)} \right) ,
\end{equation}
where $\bar{B}$ is the complement of $B$.

If $\epsilon_n= c_0 n^{-2-\delta}$ for small $\delta>0$, then $\Pr(A) \to 1$ as $n \to \infty$:
\begin{align}
    1 &\ge \Pr(A)=\Pr \left( \min\{Y_{n:2}^{*} - Y_{n:1}^{*}, \ldots, Y_{n:n}^{*} - Y_{n:n-1}^{*} \} 
    > \epsilon_n \cdot  2\normbig{\vecf{x}^{*} }  \right) \nonumber \\
    & = 1 -  \Pr \left( \min\{Y_{n:2}^{*} - Y_{n:1}^{*}, \ldots, Y_{n:n}^{*} - Y_{n:n-1}^{*} \} 
    \le \epsilon_n \cdot  2\normbig{\vecf{x}^{*} }  \right) \nonumber \\
    & \ge 1 - \left\{1- \left[ 1-(n+1) \frac{ \epsilon_n \cdot  2\normbig{\vecf{x}^{*} } }{L} \right]^n \right\} \quad\textrm{by \cref{lemma:individual_dim}
    }   \nonumber \\
    & = \left[ 1-(n+1)\epsilon_n C  \right]^n, \ \textrm{ for some positive constant } C= 2\normbig{\vecf{x}^{*} } /L, \nonumber \\
    & =  \left[ 1-(n+1)n^{-2-\delta} C_1  \right]^n \quad\textrm{for constant }C_1\equiv c_0 C \nonumber \\
    & \to 1
    \textrm{ as } n \to \infty
    \textrm{ by \cref{prop16:facts}.}
\end{align}

Consider the time dimension. 
Let $ \hat{\vecf{\beta}}_T(U_i) $ denote the TS-OLS estimator computed using its $T$ time periods data for individual $i$. 
From \cref{A8:TSOLS:CON},
we know $\forall \ \epsilon >0$, 
with probability approaching 1, 
\begin{align}
    \absbig{ \hat{Y}_{i}^{*}  -Y_i^{*}  } 
    &= \absbig{ \vecf{x}^{*\prime} \hat{\vecf{\beta}}_T(U_{i}) -\vecf{x}^{*\prime} \vecf{\beta}(U_{i})  } \nonumber \\
    &= \absbig{ \vecf{x}^{*\prime} \left(\hat{\vecf{\beta}}_T(U_{i}) - \vecf{\beta}(U_{i}) \right)  }  \nonumber \\
    &\le \normbig{ \vecf{x}^{*} } \  \normbig{\hat{\vecf{\beta}}_T(U_{i}) - \vecf{\beta}(U_{i}) } \ \textrm{by the Cauchy}-\textrm{Schwarz inequality }  \label{algebra4:YhatYBound} \\
    & <  \normbig{ \vecf{x}^{*} } \epsilon , 
\end{align}
where
$\normbig{\cdot}$ denotes the $L_2$ norm. 
That is, for any vector $\vecf{a}=(a_1, \ldots, a_K)' \in \R^K$, $\normbig{\vecf{a}}= \left( \sum_{i=1}^K a_i^2 \right)^{1/2}  $.

Notice 
\begin{align*}
    \Pr(B) & = \Pr\left( \normbig{\vecf{x}^{*}} c_0 n^{-2-\delta}>\max_{1\leq i \leq n } \absbig{ \hat{Y}_{i}^{*}  -Y_i^{*}  }  \right) \nonumber \\
    & \ge \Pr\left( \normbig{\vecf{x}^{*}} c_0 n^{-2-\delta}> \normbig{\vecf{x}^{*}} \max_{1\leq i \leq n }  \normbig{\hat{\vecf{\beta}}_T(U_{i}) - \vecf{\beta}(U_{i})}  \right)\quad
    \textrm{by \cref{algebra4:YhatYBound}} 
    \nonumber \\
    & = \Pr\left(  
    c_0 n^{-2-\delta}> \max_{1\leq i \leq n }  \normbig{\hat{\vecf{\beta}}_T(U_{i}) - \vecf{\beta}(U_{i})}  
    \right)
\end{align*}
It remains to show $\Pr(B) \to 1$ as $n,T \to \infty$ under the $n/T$ rate relation.
It is sufficient to show $\max_{1\le i \le n } \norm{\hat{\vecf{\beta}}_T(U_{i}) - \vecf{\beta}(U_{i})}  =o_p( n^{-2-\delta} )$ as $n,T \to \infty$ under the $n/T$ rate relation.

By \cref{A10:TS-OLS-finiteMean} and the weak law of large numbers, 
\begin{align*}
     \frac{1}{n}  \sum_{i=1}^n \norm{\hat{\vecf{\beta}}_T(U_{i}) - \vecf{\beta}(U_{i})}
        & = T^{-\kappa} \left( 
        \frac{1}{n} \sum_{i=1}^n T^{\kappa}  \norm{\hat{\vecf{\beta}}_T(U_{i}) - \vecf{\beta}(U_{i})}
        \right)\\ 
        & = T^{-\kappa} \left[
        \E \left( T^{\kappa} \norm{\hat{\vecf{\beta}}_T(U_{i}) - \vecf{\beta}(U_{i})}  \right) + o_p(1) 
        \right]
        = O_p \left( T^{-\kappa} \right).
\end{align*}
Thus, 
\begin{equation*}
    \max_{1\le i \le n}  \norm{\hat{\vecf{\beta}}_T(U_{i}) - \vecf{\beta}(U_{i})}  
    \le \sum_{i=1}^n  \norm{\hat{\vecf{\beta}}_T(U_{i}) - \vecf{\beta}(U_{i})}
    = n \frac{1}{n} \sum_{i=1}^n \norm{\hat{\vecf{\beta}}_T(U_{i}) - \vecf{\beta}(U_{i})}
    = n O_p( T^{-\kappa} ) 
    = O_p(n T^{-\kappa} ).
\end{equation*}

The $n/T$ rate relation $n=o\left( T^{\kappa/(3+\delta)} \right)$ implies $n^{3+\delta}/T^\kappa \to 0$ as $n,T \to \infty$, 
or equivalently $T^{-\kappa}=o( n^{-3-\delta} )$ as $n, T\to \infty$.
Therefore, 
$\max_{1\le i \le n } \norm{\hat{\vecf{\beta}}_T(U_{i}) - \vecf{\beta}(U_{i})} =O_p(n T^{-\kappa} ) = o_p ( n^{-2-\delta} )$.
\end{proof}

\Cref{lemma:individual_dim} proves a property which is used in the proof of \cref{prop4:betanorm-nrate} and \cref{thm5:largeTNconsis}.

\begin{lemma}\label{lemma:individual_dim}
    Suppose \cref{A1:iidSampling,A2:rank,A3:monoto,A4:cts:weakest,A6:idioV,A7:beta:cts:uniform,A8:TSOLS:CON,A9:LowerBoundBetaDeriv,A10:TS-OLS-finiteMean} hold.
    Then, 
    where $L$ is the lower bound of derivative in \cref{A9:LowerBoundBetaDeriv},
\begin{align*}
     \Pr\left(  \min\{Y_{n:2}^{*} - Y_{n:1}^{*}, \ldots, Y_{n:n}^{*}- Y_{n:n-1}^{*} \} \le x  \right) 
    & \le 1- [1-(n+1)x/L]^n 
    \text{ for }x \in \left[0, \frac{L}{n+1} \right]
,\\
\Pr\left(  \min\{Y_{n:2}^{*} - Y_{n:1}^{*}, \ldots, Y_{n:n}^{*}- Y_{n:n-1}^{*} \} \le x  \right) 
&=0\text{ for }x \le 0
.
\end{align*}
\end{lemma}

\begin{proof}[Proof of \cref{lemma:individual_dim} ]

Consider an element $Y_{n:k+1}^{*} - Y_{n:k}^{*} $, $ k=1, \ldots, n-1$. It is a random variable.
\begin{equation}
    Y_{n:k+1}^{*}- Y_{n:k}^{*}
=\vecf{x}^{*\prime} \left( \vecf{\beta}(U_{n:k+1}) - \vecf{\beta}(U_{n:k}) \right) , 
k=1, \ldots, n-1.
\end{equation}
Take the first order Taylor expansion 
\begin{equation} 
    \vecf{\beta}(U_{n:k+1}) - \vecf{\beta}(U_{n:k}) = 
    \vecf{\beta}'(\tilde{u})
    \underbrace{( U_{n:k+1}- U_{n:k}) }_{\sim \BetaDist(1,n) }, 
\end{equation}
where $\vecf{\beta}'(\tilde{u})$ is a $K \times 1$ vector for some $0 < \tilde{u} < 1$.
Since $( U_{n:k+1}- U_{n:k})\pconv 0$, 
each random variable $ (Y_{n:k+1}^{*}- Y_{n:k}^{*}) \pconv 0$, as $n \to \infty$.
Under \cref{A9:LowerBoundBetaDeriv}, 
$ \vecf{x}^{*\prime}\vecf{\beta}'(\tilde{u})$ has a positive lower bound $L$. 
Therefore, 
\begin{align}
     Y_{n:k+1}^{*} - Y_{n:k}^{*} 
& = \vecf{x}^{*\prime} \left( \vecf{\beta}(U_{n:k+1}) - \vecf{\beta}(U_{n:k}) \right) \nonumber \\
& = \underbrace{ \vecf{x}^{*\prime} \vecf{\beta}'(\tilde{u})}_{ \ge L  \ \textrm{by} \cref{A9:LowerBoundBetaDeriv} } ( U_{n:k+1}- U_{n:k})  \nonumber \\
& \ge L ( U_{n:k+1}- U_{n:k}).
\end{align}
Furthermore, 
\begin{align}
    \Pr\left(  Y_{n:k+1}^{*}- Y_{n:k}^{*} \le x \right) 
    & \le \Pr \left( L(U_{n:k+1}- U_{n:k}) \le x \right)  
     =    \Pr \left( U_{n:k+1}- U_{n:k} \le \frac{x}{L} \right) ,
\end{align}
and
\begin{align*}
   \Pr\left(  \min\{Y_{n:2}^{*} - Y_{n:1}^{*}, \ldots, Y_{n:n}^{*}- Y_{n:n-1}^{*} \} \le x \right)
    &= \Pr\left( \cup_{k=1}^{n-1} \{Y_{n:k+1}^{*} - Y_{n:k}^{*}  \le x \} \right)  \nonumber \\
    &\le \Pr\left( \cup_{k=1}^{n-1} \left\{ U_{n:k+1}- U_{n:k} \le \frac{x}{L} \right\} \right)  \nonumber \\
    &=
    \Pr  \left(  \min\{U_{n:2}- U_{n:1}, \ldots, U_{n:n}- U_{n:n-1}\} \le \frac{x}{L} \right).
\end{align*}
The order of convergence for random variable $\min\{ Y_{n:2}^{*}-Y_{n:1}^{*}, \ldots,Y_{n:n}^{*}-Y_{n:n-1}^{*}  \} $
is bounded by the order of convergence for random variable 
$\min\{ U_{n:2}-U_{n:1}, \ldots, U_{n:n} - U_{n:n-1}  \} $ .

Let $D_k=U_{n:k+1} -U_{n:k}$, $k=1,\ldots, n-1$. 
Let $D_0=U_{n:1}$ and $D_n=1-U_{n:n}$.
We know from \citet[pp.\ 236--238]{Wilks1962}
that 
$D_0, \ldots, D_n$ jointly follows the Dirichlet distribution with $n+1$ parameters 1. 
\begin{equation*}
    (D_0, \ldots, D_n) \sim \DirDist \underbrace{(1,\ldots, 1)}_{n+1} .
\end{equation*}
Then
we can use the distribution function of $\min\{D_0, \ldots, D_{n} \} $ 
as the upper bound 
for the distribution function of the random variable $\min\{D_1, \ldots, D_{n-1} \} $ 
\begin{equation}
\Pr \left(\min\{D_1, \ldots, D_{n-1} \} \le x \right) 
     \le \Pr \left(\min\{D_0, \ldots, D_{n} \} \le x \right) .
\end{equation}
We can derive the distribution function of $\min\{D_0, \ldots, D_{n} \} $,
that for any $x \in \left[0, \frac{1}{n+1} \right]$, 
\begin{align}\label{algebra1:CDFminD}
    \Pr \left(\min\{D_0, \ldots, D_{n} \} \le x \right) 
    & = 1 - \Pr \left(D_0 >x, \textrm{ and } \ldots, \textrm{ and }   D_{n}> x \right) 
    \nonumber \\
    & = 1 - [1-(n+1)x]^{n} .
\end{align}
$\Pr \left(\min\{D_0, \ldots, D_{n} \} \le x \right) =0 $, for $x \le 0$; 
$\Pr \left(\min\{D_0, \ldots, D_{n} \} \le x \right) =1 $, for $x \ge \frac{1}{n+1}$.

The last equality in \cref{algebra1:CDFminD} is based on the fact that 
\begin{align}\label{algebra2:}
  &  \Pr \left(D_0 >x, \textrm{ and } \ldots, \textrm{ and }   D_{n}> x \right)  \nonumber \\*
  &  = \varint[.6]_{x}^{1-nx\!} \varint[.6]_{\!x}^{1-(n-1)x-d_0\!} \varint[.6]_{x}^{1-(n-2)x-d_0-d_1\!} \!\!\!\!\cdots \! \varint[.6]_{x}^{1-x-d_0-d_1-\ldots-d_{n-2}\!} 
 \! \notag
\\* &\qquad\qquad
  f_{\DirDist} \left(d_0, d_1, \ldots, d_{n-1}, 1-\sum_{k=0}^{n-1} d_k \right)
   \,d d_{n-1} \cdots \,d d_0  \nonumber \\
  &  = \int_{x}^{1-nx} \int_{x}^{1-(n-1)x-d_0} \int_{x}^{1-(n-2)x-d_0-d_1} \cdots \int_{x}^{1-x-d_0-d_1-\ldots-d_{n-2}} 
  n! \  \,d d_{n-1} \cdots \,d d_0  \nonumber \\
  &  = n! \int_{x}^{1-nx} \int_{x}^{1-(n-1)x-d_0} \int_{x}^{1-(n-2)x-d_0-d_1} \cdots \int_{x}^{1-x-d_0-d_1-\ldots-d_{n-2}} \ 
  1\  \,d d_{n-1}  \cdots \,d d_0  \nonumber \\
  &= n! \frac{[1-(n+1)x]^n}{n!} = [1-(n+1)x]^n .
\end{align}
where $f_{\DirDist} \left(d_0, d_1, \ldots, d_{n-1}, 1-\sum_{k=0}^{n-1} d_k \right)$ 
denote the Dirichlet density function with $(n+1)$ parameter $(1,\ldots, 1)$.
\begin{equation}
    f_{\DirDist} 
    \left( 
    d_0, d_1, \ldots, d_{n-1}, 1-\sum_{k=0}^{n-1} d_k; \underbrace{1,\ldots, 1}_{n+1} 
    \right)
      = \frac{ \left[ \prod_{k=0}^{n-1} d_k^{1-1} \right]  (1-\sum_{k=0}^{n-1} d_k)^{1-1} }{ \left[ \frac{\prod_{k=0}^{n} \Gamma(1) }{\Gamma(n+1) } \right] } 
      = \Gamma(n+1) =n! ,
\end{equation}
and the gamma function with any positive integer $n$ is $\Gamma(n)=(n-1)!$.

The second to last equality in \cref{algebra2:} is based on a calculus derivation using changing of variables: 
\begin{align*}
  &  \int_{x}^{1-nx} \int_{x}^{1-(n-1)x-d_0\!\!} \int_{x}^{1-(n-2)x-d_0-d_1\!\!} \cdots\!\! \int_{x}^{1-2x-d_0-d_1-\ldots-d_{n-3}\!\!}  \int_{x}^{1-x-d_0-d_1-\ldots-d_{n-2}\!\!} \ 
  1\  \,d d_{n-1} \,d d_{n-2} \cdots \,d d_0  \nonumber \\
  & = \int_{x}^{1-nx} \int_{x}^{1-(n-1)x-d_0\!\!} \int_{x}^{1-(n-2)x-d_0-d_1\!\!} \cdots\!\! \nonumber \\
  & \qquad \qquad \int_{x}^{1-3x-d_0-d_1-\ldots-d_{n-4}\!\!}  \int_{x}^{1-2x-d_0-d_1-\ldots-d_{n-3}\!\!} \ 
  (1\!-\!2x\!-\!d_0\!-\!d_1\!-\!\ldots\!-\!d_{n-2}\!)\  \,d d_{n-2} \,d d_{n-3} \cdots \,d d_0  \nonumber \\
  & \stackrel{\text{Let } s_2=1-2x-d_0-d_1-\ldots-d_{n-2}}{=} -
  \int_{x}^{1-nx} \int_{x}^{1-(n-1)x-d_0\!\!} \int_{x}^{1-(n-2)x-d_0-d_1\!\!} \cdots 
  \nonumber \\
  & \qquad \qquad \qquad \qquad \qquad \qquad \qquad \qquad
  \int_{x}^{1-3x-d_0-d_1-\ldots-d_{n-4}\!\!}  \int_{1-3x-d_0-d_1-\ldots-d_{n-3}\!\!}^{0} \ 
  s_2 \  \,d s_2 \,d d_{n-3} \cdots \,d d_0  \nonumber \\
  & = \frac{1}{2}  \int_{x\!}^{1-nx\!\!} \int_{x\!}^{1-(n-1)x-d_0\!\!} \int_{x}^{1-(n-2)x-d_0-d_1\!\!} {\!\!\!\cdots \!\!}
   \int_{x}^{\!1-3x-d_0-d_1-\ldots-d_{n-4}\!\!} (1\!-\!3x\!-\!d_0\!-\!d_1\!-\!\ldots\!-\!d_{n-3})^2 
   \,d d_{n-3} \cdots \,d d_0  \nonumber \\
  & \stackrel{\text{Let } s_3=1-3x-d_0-d_1-\ldots-d_{n-3}}{=} 
 - \frac{1}{2}
  \int_{x\!}^{1-nx\!\!} \int_{x\!}^{1-(n-1)x-d_0\!\!} \int_{x\!}^{1-(n-2)x-d_0-d_1\!\!} {\!\!\cdots\!\!} 
\int_{1-4x-d_0-d_1-\ldots-d_{n-4}\!\!}^{0}   s_3^2 
  \,d s_3 \,d d_{n-4} \cdots \,d d_0 
  \nonumber \\
  & =   \frac{1}{2\!\cdot\! 3}\!  \int_{x\!}^{1-nx\!\!} \int_{x\!}^{1-(n-1)x-d_0\!\!} \int_{x\!}^{1-(n-2)x-d_0-d_1\!\!} {\!\!\!\cdots \!\!}
  \int_{x}^{1-4x-d_0-d_1-\ldots-d_{n-5}\!\!}   (\!1\!-\!4x\!-\!d_0\!-\!d_1\!-\!\ldots\!-\!d_{n-4})^3 
   \,d d_{n-4} \cdots \,d d_0 \nonumber \\
  & = \ldots \nonumber \\
  & = \frac{1}{(n-1)!} \int_{x}^{1-nx} (1-nx-d_0)^{n-1} \,d d_0 \nonumber \\
  & \stackrel{\text{Let } s=1-nx-d_0}{=}  -\frac{1}{(n-1)!} \int_{1-(n+1)x}^{0} s^{n-1}  \,d s \nonumber \\
  & = \frac{1}{n!} \left. s^n \right\vert_{0}^{1-(n+1)x} \nonumber \\
  &= \frac{[1-(n+1)x]^n}{n!} .
\end{align*}

For the special case $x=0$,
\begin{multline*}
    \int_{0}^{1}\! \int_{0}^{1-d_0} \!\! \int_{0}^{1-d_0-d_1\!} \!\!\!\!\cdots \! \int_0^{1-d_0-d_1-\ldots-d_{n-2}\!} 
 \! f_{\DirDist} \left(d_0, d_1, \ldots, d_{n-1}, 1-\sum_{k=0}^{n-1} d_k \right)
   \,d d_{n-1} \cdots \,d d_0 
\\
= n! \frac{1}{n!}
= 1 .
\end{multline*}

Altogether, we can derive the upper bound for the distribution function of the random variable 
$\min\{Y_{n:2}^{*} - Y_{n:1}^{*}, \ldots, Y_{n:n}^{*}- Y_{n:n-1}^{*} \}$. 
For $x \in \left[0, \frac{L}{n+1} \right]$, 
\begin{align}
     \Pr\left(  \min\{Y_{n:2}^{*} - Y_{n:1}^{*}, \ldots, Y_{n:n}^{*}- Y_{n:n-1}^{*} \} \le x  \right) 
    & \le    \Pr\left(  \min\{U_{n:2}- U_{n:1}, \ldots, U_{n:n}- U_{n:n-1}\} \le \frac{x}{L} \right) \nonumber \\
    & \le 1- [1-(n+1)x/L]^n    .
\label{eqn:P-min-Ystar-diff}
\end{align}

\end{proof}

\begin{proof}[Proof of \cref{thm5:largeTNconsis}]

\textbf{Main idea of proof}

To obtain the consistency of the coefficient estimator at any rank in the large-$T$ large-$n$ 
model \cref{T>K:idio-out}, 
it requires to ensure two things.
One is that 
in the fixed-$n$ large-$T$ setting, 
the estimated $n$ individual coefficients should be consistent via its time-dimension observations. 
The other is that the ordering of the predicted outcome values $\hat{Y}^*$ should correctly represent the ordering of the unobserved rank variable values.

The first requirement is met by \Cref{A8:TSOLS:CON}. 
The second requirement is a non-trivial work to show.
Let $Y^{*}$ denote the 
true
outcome values at $\vecf{X}=\vecf{x}^{*}$ 
computed using the unknown true coefficient. 
Let $\hat{Y}^{*}$ denote the fitted outcome values 
at $\vecf{X}=\vecf{x}^{*}$ 
computed using the estimated individual coefficients. 
\Cref{A3:monoto} indicates  
the ordering of outcome values $Y^{*}$ can represent the ordering of unobserved rank variable $U$.
However,
we can only obtain the fitted outcome values $\hat{Y}^{*}$ instead of the true outcome values $Y^{*}$ in model \cref{T>K:idio-out}.
Since the estimated individual coefficient has estimation error,
the ordering of the fitted outcome values $\hat{Y}^{*}$ might be different from the ordering of true outcome values $Y^{*}$.
If so, the ordering of $\hat{Y}^{*}$ cannot represent the ordering of rank variable $U$.

The idea is to show the fitted outcome value $\hat{Y}^{*} $ is close enough to 
the true outcome value $Y^{*}$.
Technically, 
we hope to show the fitted outcome value $\hat{Y}^{*} $ is within some $\epsilon$-ball of the true value $Y^{*}$, 
with radius less than the smallest value of all $Y^{*}_{n:k+1}-Y^{*}_{n:k}$, $k=1, \ldots, n-1$. 
The $\epsilon$-ball between $\hat{Y}^{*} $ and $Y^{*}$
measures the time-dimension estimation error 
and it is related with the convergence rate of $T$.
The $\{ Y^{*}_{n:k+1}-Y^{*}_{n:k} \}_{k=1}^{n-1}$'s are random variables with respect to the individual dimension $n$ 
and they are related with the convergence rate of $n$.
We hope to find certain rate relation between $T$ and $n$, at which 
the $\epsilon $-ball shrinks faster than 
the smallest value of all $Y^{*}_{n:k+1}-Y^{*}_{n:k}$, $k=1, \ldots, n-1$ does
in order to guarantee the 
order of fitted outcome values $\hat{Y}^{*}$ represent
the order of the true outcome values $Y^{*}$ 
and the order of the unobserved rank variable values $U$.

The time dimension is considered in the proof of \cref{prop4:betanorm-nrate}.
The individual dimension is considered in \cref{lemma:individual_dim}.

\noindent \textbf{Correct ordering under rate relation}

\begin{tikzpicture}
\path (-3,0) node(x) {} 
      (12,0) node(y) {};
\draw (x) -- (y);
\filldraw 
(-2.5,0)  node[align=center] { ( } --
(-1.5,0) circle (0.5pt) node[align=center,   above] { $\hat{Y}_{i}^{*} $ } --
(-0.5,0) circle (0.5pt) node[align=center,   above] {$(\stackrel{ ?}{=}  \hat{Y}_{n:1}^{*}) $ \\  } --
(-1,0) circle (2pt) node[align=center,   below] {\\$Y_{n:1}^{*}$ \\ $(=Y^{*}_i)$ } --
(0.5,0)  node[align=center] { ) } --
(1.5,0)  node[align=center] { ( } --
(3,0) circle (2pt) node[align=center,   below] {\\$Y_{n:2}^{*}$} --
(4.5,0)  node[align=center] { ) } --
(7,0)  node[align=center, below] {\\$\cdots$}  -- 
(9,0) circle (2pt) node[align=center,  below] {\\$Y_{n:n}^{*}$ };
\end{tikzpicture}

\Cref{prop4:betanorm-nrate} implies that
when $n$ and $T$ are sufficiently large with the rate relation 
$ n=o( T^{\kappa/(3+\delta)} ) $,
the order of fitted outcome values $\hat{Y}^{*}$ can represent the order of the true outcome values $Y^{*}$, 
which further represents
the order of the unobserved rank variable values $U$.
Therefore, 
\begin{equation}\label{appdx:4:correct-ordering}
    \hat{Y}_{n:k}^{*}= \vecf{x}^{*\prime} \hat{\vecf{\beta}} ( U_{n:k} ), k=1, \ldots, n. 
\end{equation}
The order of the fitted outcome values indicates the order of the rank variable values.
Therefore, 
the estimator $\hat{\vecf{\beta}}( \tau)$,
defined as the coefficient estimates associated with the $\lceil n\tau \rceil$th order statistic of the fitted outcome values, 
is actually the coefficient estimates for the person with the $\lceil n\tau \rceil$th order statistic of rank variable,
\begin{equation}\label{appx:def:betahat}
    \hat{\vecf{\beta}} \left( \tau \right) =  \hat{ \vecf{\beta}} \left( U_{n:\lceil n\tau \rceil  } \right) .
\end{equation}
Note that \cref{prop4:betanorm-nrate} guarantees the occurrence of \cref{appdx:4:correct-ordering} and thus \cref{appx:def:betahat} with probability approaching one.

\paragraph{Consistency}

Next we are going to show the estimator is uniform consistent in probability under the $n/T$ rate relation.
That is to show, 
as $n  \to \infty, T \to \infty $, and $n=o\left(T^{\frac{\kappa}{3+\delta}}\right)$, 
for any $\eta >0$, 
\begin{align}
    \Pr \left(  \sup_{\tau \in (0,1)} 
    \normbig{ \hat{\vecf{\beta}}( \tau) - \vecf{\beta}( \tau) } 
    >\eta
  \right) \to 0.
\end{align}
Let $ A$ denote the event which will result in the correct ordering. 
(It is the same as event $W$ previously.) 
\begin{align}
    A &= \left\{ 
\frac{\min\{Y_{n:2}^{*} - Y_{n:1}^{*}, \ldots, Y_{n:n}^{*} - Y_{n:n-1}^{*} \} }{2}>  \absbig{ \hat{Y}_{T,1}^{*} - Y_{1}^{*}} , 
\textrm{ and } 
\ldots, 
\right. \nonumber \\
        & \quad \left. \ldots,  \textrm{ and } 
        \frac{\min\{Y_{n:2}^{*} - Y_{n:1}^{*}, \ldots, Y_{n:n}^{*} - Y_{n:n-1}^{*} \} }{2}>  \absbig{ \hat{Y}_{T,n}^{*} - Y_{n}^{*}}  
        \right\} \nonumber \\
&= \left\{ 
        \frac{\min\{Y_{n:2}^{*} - Y_{n:1}^{*}, \ldots, Y_{n:n}^{*} - Y_{n:n-1}^{*} \} }{2}> \max_{1\le i \le n} \absbig{ \hat{Y}_{i}^{*} - Y_{i}^{*}} 
        \right\} .
\end{align}
From 
\cref{prop4:betanorm-nrate},
we know under the $n/T$ rate relation, 
the event $A$ eventually occurs with probability 1,
i.e., 
as $n, T \to \infty $, with $n=o\left(T^{\frac{\kappa}{3+\delta}}\right)$, 
\begin{equation}
    \Pr \left( A \right) \to 1.
\end{equation}
Let $\bar{A}$ denote the complement of event $A$. 
Then $\Pr \left(\bar{A}\right) \to 0, $
as $n, T \to \infty $ with $n=o\left(T^{\frac{\kappa}{3+\delta}}\right)$.

For any $\eta >0$, 
\begin{align}\label{algebra6:}
     \Pr \left(  \sup_{\tau \in (0,1)} 
    \normbig{ \hat{\vecf{\beta}}( \tau) - \vecf{\beta}( \tau) } 
    >\eta
  \right) 
  & \le 
 \underbrace{ \Pr \left(\bar{A}\right) }_{\to 0 }
 + 
 \underbrace{ \Pr \left( A \right) }_{\to 1 }
 \Pr \left( \sup_{\tau \in (0,1)} 
    \normbig{ \hat{\vecf{\beta}}( \tau) - \vecf{\beta}( \tau) } 
    >\eta \mid A \right) .
\end{align}
In \cref{appdx:4:correct-ordering} we showed 
under event $A$, 
the order of fitted outcome values $\hat{Y}^{*}$ correctly represents the order of unobserved rank variable. 
Then 
\begin{align}
 0 &\le   \Pr \left( \sup_{\tau \in (0,1)} 
    \normbig{ \hat{\vecf{\beta}}( \tau) - \vecf{\beta}( \tau) } 
    >\eta \mid A \right) \nonumber \\
   & =      \Pr \left( \sup_{\tau \in (0,1)} 
     \normbig{  \hat{\vecf{\beta}} \left( U_{n:\lceil \tau n \rceil  } \right) - \vecf{\beta}( \tau)}
    >\eta \mid A \right) \nonumber \\
    & =     \Pr \left( \sup_{\tau \in (0,1)} 
     \normbig{  \hat{\vecf{\beta}} \left( U_{n:\lceil \tau n \rceil  } \right) - \vecf{\beta} \left( U_{n:\lceil \tau n \rceil  } \right)
    + \vecf{\beta} \left( U_{n:\lceil \tau n \rceil  } \right) -
    \vecf{\beta}( \tau)}
    >\eta \mid A \right) \nonumber \\
    & \le \Pr \left( \sup_{\tau \in (0,1)} 
     \normbig{  \hat{\vecf{\beta}} \left( U_{n:\lceil \tau n \rceil  } \right) - \vecf{\beta} \left( U_{n:\lceil \tau n \rceil  } \right) } 
    +  \sup_{\tau \in (0,1)}  \normbig{ \vecf{\beta} \left( U_{n:\lceil \tau n \rceil  } \right) -
    \vecf{\beta}( \tau)}
    >\eta \mid A \right) \nonumber \\
    & = \Pr \left(  
     \sup_{\tau \in (0,1)}  \normbig{ \vecf{\beta} \left( U_{n:\lceil \tau n \rceil  } \right) -
    \vecf{\beta}( \tau)}
    >\eta 
    - \sup_{\tau \in (0,1)} 
     \normbig{  \hat{\vecf{\beta}} \left( U_{n:\lceil \tau n \rceil  } \right) - \vecf{\beta} \left( U_{n:\lceil \tau n \rceil  } \right) }
     \mid A \right) \nonumber \\
    & = 
    \Pr \left( \sup_{\tau \in (0,1)} 
     \normbig{  \hat{\vecf{\beta}} \left( U_{n:\lceil \tau n \rceil  } \right) - 
     \vecf{\beta} \left( U_{n:\lceil \tau n \rceil  } \right) }
     \ge \eta \mid A
     \right) \nonumber \\
     & \quad + \Pr \left( 
     \sup_{\tau \in (0,1)} 
     \normbig{  \hat{\vecf{\beta}} \left( U_{n:\lceil \tau n \rceil  } \right) - 
     \vecf{\beta} \left( U_{n:\lceil \tau n \rceil  } \right) }
     < \eta \ \textrm{ and } \right. \nonumber \\
     &\qquad \qquad  \left. \sup_{\tau \in (0,1)}  \normbig{ \vecf{\beta} \left( U_{n:\lceil \tau n \rceil  } \right) -
    \vecf{\beta}( \tau)}
    >\eta
    - \sup_{\tau \in (0,1)} 
     \normbig{  \hat{\vecf{\beta}} \left( U_{n:\lceil \tau n \rceil  } \right) - 
     \vecf{\beta} \left( U_{n:\lceil \tau n \rceil  } \right) }
    \mid A \right) \nonumber \\    
    & = 
    \Pr \left( \sup_{\tau \in (0,1)} 
     \normbig{  \hat{\vecf{\beta}} \left( U_{n:\lceil \tau n \rceil  } \right) - 
     \vecf{\beta} \left( U_{n:\lceil \tau n \rceil  } \right) }
     \ge \eta \mid A
     \right)
     \nonumber \\
     & \quad +
     \Pr \left( 
      \sup_{\tau \in (0,1)}  \normbig{ \vecf{\beta} \left( U_{n:\lceil \tau n \rceil  } \right) -
    \vecf{\beta}( \tau)}
    >\eta
    - \sup_{\tau \in (0,1)} 
     \normbig{  \hat{\vecf{\beta}} \left( U_{n:\lceil \tau n \rceil  } \right) - 
     \vecf{\beta} \left( U_{n:\lceil \tau n \rceil  } \right) } >0
    \mid A \right) .
\end{align}
Let $K$ denote the event $K= \left\{ \sup_{\tau \in (0,1)} 
     \normbig{  \hat{\vecf{\beta}} \left( U_{n:\lceil \tau n \rceil  } \right) - 
     \vecf{\beta} \left( U_{n:\lceil \tau n \rceil  } \right) }
     \ge \eta 
     \right\}$.
\cref{A8:TSOLS:CON} \cref{A8_uni_consist} assumes $\Pr(K) \to 0$, as $n\to \infty, T \to \infty$.
Then,
\begin{equation}\label{prob-KcondiA}
 \Pr(K \mid A)= \frac{\Pr(K \textrm{ and } A) }{ \Pr(A) } 
    \le \frac{ \overbrace{\Pr(K)}^{\to 0}  }{ \underbrace{\Pr(A)}_{\to 1} } 
    \to 0, \textrm{ as } n\to \infty, T \to \infty,
    \textrm{ with } n=o\left(T^{\frac{\kappa}{3+\delta}}\right) .
\end{equation}
Let $\gamma=\eta
    - \sup_{\tau \in (0,1)} 
     \normbig{  \hat{\vecf{\beta}} \left( U_{n:\lceil \tau n \rceil  } \right) - 
     \vecf{\beta} \left( U_{n:\lceil \tau n \rceil  } \right) }>0$.
Denote the event
\begin{equation*}
    E= \left\{        \sup_{\tau \in (0,1)}  \normbig{ \vecf{\beta} \left( U_{n:\lceil \tau n \rceil  } \right) -
    \vecf{\beta}( \tau)}
    >
    \gamma 
    \right\}. 
\end{equation*}
From \cref{lem:simplied:uniform:consistency}, we know
\begin{equation}
    \Pr\left( E \right) \to
    0, \ \textrm{ as } n \to \infty .
\end{equation}
The equality 
$    \Pr\left( E \right) = 
    \Pr\left( E \mid A \right) \Pr\left( A \right) + \Pr\left( E \mid \bar{A} \right) \Pr \left( \bar{A} \right)
$
implies 
\begin{equation}
    \Pr\left( E \mid A \right)= \frac{\Pr\left( E \right)- \Pr\left( E \mid \bar{A} \right) \Pr \left( \bar{A} \right) }{\Pr\left( A \right)}.
\end{equation}
Since $\Pr(E) \to 0$, $ \Pr\left( A \right) \to 1$, and $\Pr \left( \bar{A} \right) \to 0$ under the $n/T$ rate relation, 
\begin{equation}\label{prob-EcondiA}
    \Pr\left( E \mid A \right) 
    \to 0, 
\ \textrm{ as } n \to \infty, T \to \infty, \textrm{ with } n=o\left(T^{\frac{\kappa}{3+\delta}}\right) .
\end{equation}
Therefore, 
\begin{align}
 0 &\le   \Pr \left( \sup_{\tau \in (0,1)} 
    \normbig{ \hat{\vecf{\beta}}( \tau) - \vecf{\beta}( \tau) } 
    >\eta \mid A \right) \nonumber \\
    & \le  1 \cdot
    \underbrace{ \Pr \left( \sup_{\tau \in (0,1)} 
     \normbig{  \hat{\vecf{\beta}} \left( U_{n:\lceil \tau n \rceil  } \right) - 
     \vecf{\beta} \left( U_{n:\lceil \tau n \rceil  } \right) }
     \ge \eta \mid A
     \right)
     }_{ = \Pr\left( K\mid A\right)  \to 0, \textrm{ as } n \to \infty , T \to \infty, 
     \textrm{ with } n=o\left(T^{\frac{\kappa}{3+\delta}}\right)
     \textrm{ by } \cref{prob-KcondiA} } \nonumber \\
     & \quad +
     \underbrace{\Pr \left( 
      \sup_{\tau \in (0,1)}  \normbig{ \vecf{\beta} \left( U_{n:\lceil \tau n \rceil  } \right) -
    \vecf{\beta}( \tau)}
    >\eta
    - \sup_{\tau \in (0,1)} 
     \normbig{  \hat{\vecf{\beta}} \left( U_{n:\lceil \tau n \rceil  } \right) - 
     \vecf{\beta} \left( U_{n:\lceil \tau n \rceil  } \right) } >0
    \mid A \right) }_{ = \Pr\left( E\mid A\right)  \to 0, \textrm{ as } n \to \infty, T \to \infty, \textrm{ with } n=o\left(T^{\frac{\kappa}{3+\delta}}\right) \textrm{ by } \cref{prob-EcondiA} 
    } \nonumber \\ & \to 0.    
\end{align}
Thus,  
\begin{equation}
    \Pr \left( \sup_{\tau \in (0,1)} 
    \normbig{ \hat{\vecf{\beta}}( \tau) - \vecf{\beta}( \tau) } 
    >\eta \mid A \right) \to 0, \textrm{ as } n \to \infty, T \to \infty, \textrm{ with } n=o\left(T^{\frac{\kappa}{3+\delta}}\right) . 
\end{equation}
And altogether in \cref{algebra6:} 
\begin{align}
     \Pr \left(  \sup_{\tau \in (0,1)} 
    \normbig{ \hat{\vecf{\beta}}( \tau) - \vecf{\beta}( \tau) } 
    >\eta
  \right) 
  & \le 
 \underbrace{ \Pr \left(\bar{A}\right) }_{\to 0 }
 + 
 \underbrace{ \Pr \left( A \right) }_{\to 1 }
 \underbrace{ \Pr \left( \sup_{\tau \in (0,1)} 
    \normbig{ \hat{\vecf{\beta}}( \tau) - \vecf{\beta}( \tau) } 
    >\eta \mid A \right)
    }_{\to 0 } \nonumber \\
    & \to 0. 
\end{align}

Therefore, 
the coefficient function estimator is uniformly consistent 
to the true coefficient function 
on $(0,1)$ 
as $n \to \infty$, $T\to \infty$, with $n$ and $T$ rate relation $n=o(T^{\kappa/(3+\delta)})$.
\end{proof}

\begin{proof}[Proof of \cref{thm13:largeTNasynormal}]

Consider the standard uniform distribution $F_U$
with domain $(s,t)=(0,1)$. 
It is continuously differentiable on $(0,1)$ 
with a 
strictly positive constant derivative $f_U=1$ on $(0,1)$. 
Therefore, 
under \cref{A12:betafn:forAsyNormal}, 
from \cref{lemma19:asyNormal},
\begin{equation}
  \sqrt{n} \left( \beta_k \circ F_{n,U}^{-1} - \beta_k \circ F_U^{-1}
    \right)
    \weaklyto 
    \beta'_k \left( F_U^{-1} \right)  \mathbb{G} 
    = 
    \beta'_k \mathbb{G} 
    , \ \textrm{in} \ \ell^{\infty}(0,1) .
\end{equation}
The last equality is due to $F_U$ is an identity function on $(0,1)$.
The limiting process is Gaussian with mean zero and covariance function 
\begin{equation*}
    \beta'_k \left( F_U^{-1}(u) \right) 
    \beta'_k \left( F_U^{-1}(v) \right) 
    (u \land v -uv )
    = \beta'_k \left( u \right) 
    \beta'_k \left( v \right) 
    (u \land v -uv )
    , \ u,v \in \R .
\end{equation*}

Under  \cref{A1:iidSampling,A2:rank,A3:monoto,A4:cts:weakest,A6:idioV,A7:beta:cts:uniform,A8:TSOLS:CON,A9:LowerBoundBetaDeriv,A10:TS-OLS-finiteMean} 
and the $n/T$ rate relation 
$n=o \left(
T^{\frac{\kappa}{3+\delta}}
\right) 
$, 
we know the ordering of rank variable can be correctly discovered,
as shown in the proof of \cref{thm5:largeTNconsis};
The individual with the $k$th ordering in the true outcome values 
$Y^{*} = \vecf{x}^{*\prime} \vecf{\beta}(U_{n:k})$
at $X=x^{*}$ 
is also the same individual 
who has the $k$th ordering in the fitted outcome values 
$\hat{Y}^{*}=\vecf{x}^{*\prime} \hat{\vecf{\beta}}(U_{n:k})$
at $X=x^{*}$.
Specifically
\Cref{A10:TS-OLS-finiteMean} implies that the gap between the estimated individual coefficient and the true coefficient in terms of $T$ order. 
\begin{equation}
         \hat{\beta}_k(U_i)- \beta_k(U_i)
         = O_p \left( T^{-\kappa} \right), 
         \ \textrm{as} \  T \to \infty , 
         \textrm{ for } \kappa>0.
\end{equation}
Therefore, the coefficient estimator can be written as 
\begin{align}
        \hat{\beta}_k \left( \tau \right)
= \hat{\beta}_k \left( U_{n:\lceil \tau n \rceil  } \right)
& = \hat{\beta}_k \left( U_{ \sigma(\lceil \tau n \rceil)  } \right) \nonumber \\
& = \beta_k \left( U_{ \sigma(\lceil \tau n \rceil)  } \right) 
+O_p \left( T^{-\kappa} \right) \nonumber \\
& = \beta_k \left( U_{ n:\lceil \tau n \rceil  } \right) 
+O_p \left( T^{-\kappa} \right).
\end{align}
Since $U_{n:\lceil \tau n \rceil  }=F_{n,U}^{-1}(\tau) $ and $\tau=F_U^{-1}(\tau)$, then
\begin{align}
  \sqrt{n} \left( \hat{\beta}_k (\cdot) - \beta_k (\cdot)
    \right) 
    & =
  \sqrt{n} \left( \beta_k \circ F_{n,U}^{-1} + O_p \left( T^{-\kappa} \right) - \beta_k \circ F_U^{-1}
    \right) 
    \nonumber \\
    & =
  \sqrt{n} \left( \beta_k \circ F_{n,U}^{-1} - \beta_k \circ F_U^{-1}
    \right) 
    + n^{1/2} O_p \left( T^{-\kappa} \right)
    \nonumber \\
    & \weaklyto 
    \beta'_k \left( F_U^{-1} \right)  \mathbb{G} 
    =   
    \beta'_k   \mathbb{G}, 
    \textrm{ in} \ \ell^{\infty}(0,1) ,
    \textrm{ as } n \to \infty, T \to \infty, 
\end{align}
since 
$n^{1/2} O_p \left( T^{-\kappa} \right) \pconv 0$
under the rate relation 
$n=o \left(
T^{\frac{\kappa}{3+\delta}}
\right) 
$. 
That is, 
$ 0 \le 
\frac{n}{ T^{2\kappa}  } \le \frac{n}{ T^{\kappa/(3+\delta)}  }
\to 0, 
    \textrm{ as } n \to \infty, T \to \infty,
$ for $\kappa>0$, $\delta>0$.\footnote{The $\delta$ here is the same as it is introduced in the proof of \cref{thm5:largeTNconsis}.}
\end{proof}

\begin{proof}[Proof of \cref{cor14:asyNormal}]
It is straightforward to apply \cref{thm13:largeTNasynormal} the specific point $\tau \in (0,1)$.
\end{proof}

\section{Additional proofs}\label{sec:appx:addproof}

\cref{prop16:facts} proves a property which is used in the proof of \cref{prop4:betanorm-nrate}.

\begin{proposition}\label{prop16:facts}
For some small positive $\delta>0$, 

(i)
\begin{equation}
    \left[1-(n+1)
    \frac{2\normbig{\vecf{x}^{*} }}{L} 
    c_0 n^{-3-\delta}
    \right]^{n^2} 
    \to 1, 
    \textrm{ as } n \to \infty; 
\end{equation}

(ii)
\begin{equation}
    \left[1-(n+1)
    \frac{2\normbig{\vecf{x}^{*} }}{L} 
    c_0 n^{-3}
    \right]^{n^2} 
    \to e^{-C}, 
    \textrm{ as } n \to \infty; 
\end{equation}

(iii)
\begin{equation}
    \left[1-(n+1)
    \frac{2\normbig{\vecf{x}^{*} }}{L} 
    c_0 n^{-3+\delta}
    \right]^{n^2} 
    \to 0,
    \textrm{ as } n \to \infty, 
\end{equation}
where $c_0$ is a constant, 
and $C=  \frac{2\normbig{\vecf{x}^{*} }}{L} 
    c_0 $ is a constant. 
\end{proposition}

\begin{proof}[Proof of \cref{prop16:facts}] We prove (i) here. The proof of (ii) and (iii) can be obtained in a similar way.
 The proof of (ii) and (iii) are identical to (i) until the last equality of computing $\lim_{n\to \infty} \ln a$, 
at which the results differ depending on $\delta>0, \delta=0$, or $\delta<0$ in the derivation of (i).

Let 
\begin{equation}
    a= \left[1-(n+1)
    \frac{2\normbig{\vecf{x}^{*} }}{L} 
    c_0 n^{-3-\delta}
    \right]^{n^2}, 
\end{equation}
    then 
    \begin{equation}
        \ln a= n^2 \ln\left[1-(n+1)
    C
    n^{-3-\delta} 
    \right] , \textrm{ for large enough $n$,}
    \end{equation}
    where $C=(2\norm{\vecf{x}^{*}}/L) 
    c_0 $ is some constant. 
    The limit of $\ln a$ is 
\begin{align*}
\lim_{n\to \infty} \ln a 
    & = 
    \lim_{n\to \infty} 
    \frac{ \ln \left[1- (n+1)C n^{-3-\delta} \right] }{ n^{-2} },
    \nonumber \\
    & = \lim_{n\to \infty} 
    \frac{- \frac{1}{1-(n+1)C n^{-3-\delta}} (-C) \left[ (2+\delta)n^{-3-\delta} + (3+\delta)n^{-4-\delta} \right]  }{-2n^{-3} }
    \ \textrm{ by L'Hospital's Rule}  \nonumber \\
    & = \lim_{n\to \infty} 
    \frac{- \frac{C}{2} \frac{1}{n^{\delta} } \left[2+\delta+\frac{3+\delta}{n} \right] }{1- \frac{C(n+1)}{n^{3+\delta}} } =0 .
\end{align*}
The last equality is because $  \frac{1}{n^{\delta} } \to 0, \ \left[2+\delta+\frac{3+\delta}{n} \right] \to 2+\delta, 
    \ \textrm{ and } \left[1- \frac{C(n+1)}{n^{3+\delta}} \right] \to 1,
$
as $n \to \infty $.
Therefore, 
by the continuous mapping theorem, 
\begin{equation*}
    \lim_{n\to \infty}  a 
    =\lim_{n\to \infty} e^{\ln a}
    =e^{\lim_{n\to \infty} \ln a}
    = e^0 =
    1.
\qedhere
\end{equation*}
\end{proof}

\begin{lemma}\label{lem:simplied:uniform:consistency}
    Under \cref{A7:beta:cts:uniform},
    for all $k=1, \ldots, K$,
\begin{equation}\label{thm:PfinBetaUniformConsis:element}
   \sup_{\tau \in (0,1)} 
   \absbig{ 
   \beta_k \left( U_{n:\lceil \tau n \rceil  } \right)-\beta_k (\tau) 
   } 
   \asconv 0, 
   \ \textrm{as }\ n \to \infty .
\end{equation}
\end{lemma}

\begin{proof}[Proof of \cref{lem:simplied:uniform:consistency}]\label{pf:consistency:uniform }

Let $F_U(\cdot)$ denote the distribution function of the standard uniform distributed random variable $U$. 
The function $F_U(\cdot)$ is continuous on $[0,1]$.
It can be written out explicitly as 
    \begin{equation}
        F_U(u) =
        \left\{ {
\begin{array}{cc}
 0,    & u \le 0   \\
 u,    & 0 \le  u \le 1 \\
 1,    &  u\ge 1 .
\end{array}
} 
\right.
    \end{equation}
Denote the $\tau$-quantile of $F_U$ as 
\begin{equation}
   Q_U(\tau) = F_U^{-1}(\tau) 
   \equiv \inf\{x:F_U(x)\ge \tau \}.
\end{equation}
Both the distribution function and the quantile function 
of the standard uniform distributed random variable $U$
are
identity functions on $(0,1)$. 
\begin{align}
    F_U(u) & = u,\ \textrm{for all}\  u \in (0,1) ;\\
Q_U(\tau) 
& = \tau, \ \textrm{for all}\ \tau \in (0,1) .
\end{align}
Thus, $F_U= I$ and $Q_U=I$, where $I$ is the identity function. 
Define the empirical distribution function of a size $n$ sample 
of the standard uniform distributed 
$U_1, \ldots, U_n$ 
as
\begin{equation}
    F_{n,U}(u)=\frac{1}{n} \sum_{i=1}^n \Ind{U_i \le u} ,
\end{equation}
where $\Ind{\cdot}$ is the indicator function.
Define the sample quantile function as the inverse of the empirical distribution function 
\begin{equation}
    \hat{Q}_U(\tau)
    \equiv  F_{n,U}^{-1}(\tau) = \inf\{x:F_{n,U}(x)\ge \tau \} .
\end{equation}
It is well known that the uniform sample quantile function is uniformly convergent. 
(See \citet[Thm.\ 3, p.\ 95]{ShorackWellner1986})
\begin{equation}
 \sup_{\tau \in (0,1)} \absbig{ F_{n,U}^{-1}(\tau) - I(\tau) } 
  \asconv 0, \ \textrm{ as } n \to \infty .
\end{equation}
Moreover, the standard uniform distributed $U$ has property that 
\begin{equation}
 F_{n,U}^{-1}(\tau )   = U_{n:\lceil \tau n\rceil} .
\end{equation}
Therefore, 
\begin{equation}\label{lemma:Uconsistent:uniform}
  \sup_{\tau \in (0,1)} \absbig{ U_{n:\lceil \tau n \rceil  } - \tau} 
  = \sup_{\tau \in (0,1)} \absbig{ F_{n,U}^{-1}(\tau) -I(\tau) } 
  \asconv 0, \ \textrm{ as } n \to \infty .
\end{equation}
That is, the sequence of functions $g_n(\tau)=U_{n:\lceil\tau n \rceil }$ uniformly converges to function $g(\tau)=\tau$ on $(0,1)$. 
Equivalently, with probability 1,
\begin{equation}
    \forall \epsilon >0, \ \exists N_{\epsilon} \in \mathbb{N} 
    \ \textrm{such that } \
    \forall n> N_{\epsilon}, 
   \ \forall \tau \in (0,1), 
  \  \absbig{U_{n:\lceil \tau n \rceil  }-\tau } <\epsilon.
\end{equation}

The uniform continuity \Cref{A7:beta:cts:uniform} states that for all $k=1, \ldots, K$, 
$\forall \epsilon > 0, \exists \  \delta_\epsilon >0, \textrm{ such\ that } \forall x, y \in (0,1), \absbig{x-y} \le \delta_\epsilon \implies \absbig{\beta_k(x)-\beta_k(y)} < \epsilon$. 
Therefore, for all $k=1, \ldots, K$, 
\begin{equation*}
    \forall \epsilon > 0,
    \exists N_{\delta_\epsilon} \in \mathbb{N} 
    \ \textrm{such that } \
    \forall n > N_{\delta_\epsilon}
    \ \textrm{and } \ 
    \forall \tau \in (0,1), 
    \ 
    \absbig{U_{n:\lceil \tau n \rceil  }-\tau }  < \delta_\epsilon,
  \textrm{with probability} \ 1,
\end{equation*}
then 
\begin{equation}
         \absbig{\beta_k \left( U_{n:\lceil \tau n \rceil  } \right) - \beta_k (\tau) } < \epsilon .
\end{equation}
The sequence of functions 
$\beta_k \left( U_{n:\lceil \tau n \rceil  } \right)$ uniformly converges almost surely to the corresponding function $\beta_k(\tau)$ as $n$ goes to infinity.
For all $k=1, \ldots, K$, $ \sup_{\tau \in (0,1)} 
   \absbig{ 
   \beta_k \left( U_{n:\lceil \tau n \rceil  } \right)-\beta_k (\tau) 
   } 
   \asconv 0, 
   \ \textrm{as }\ n \to \infty $. 
\end{proof}

Assume a general version of \cref{A12:betafn:forAsyNormal} about the differentiability of the coefficient function. 
Under this assumption, 
we build the uniform asymptotic normality of the empirical process as the coefficient function composited with any general distribution function. 
This provide the intermediate results for \cref{thm13:largeTNasynormal}.

\begin{assumption}\label{A10General:betafn:forAsyNormal}
Consider the $K$-vector coefficient function 
$\vecf{\beta}: (s,t) \subset \R \mapsto \R^{K} $.
Assume each of its element function $\beta_k: (s,t) \subset \R \mapsto \R$, $k=1, \ldots, K$, 
is differentiable with uniform continuous and bounded derivative.
\end{assumption}

\begin{lemma}[Asymptotic normality of empirical process]\label{lemma19:asyNormal}
Let $F$ be a distribution function that have compact support $[s,t]$ 
and be continuously differentiable on its support with strictly positive derivative $f$.
Let $F_n$ be the empirical distribution of an iid sample from $F$ with size $n$.
Suppose \Cref{A10General:betafn:forAsyNormal} holds.
Then the sequence of empirical processes 
$\sqrt{n} \left( \beta_k \circ F_n^{-1} - \beta_k \circ F^{-1} \right)$
converges in distribution in the space $\ell^{\infty}(0,1) $ to a limit Gaussian process
\begin{equation}\label{eqn:lemma9}
  \sqrt{n} \left( \beta_k \circ F_n^{-1} - \beta_k \circ F^{-1}
    \right)
    \weaklyto 
    \beta'_k \left( F^{-1} \right) \frac{ \mathbb{G} }{ f \left( F^{-1} \right) }, \ \textrm{in} \ \ell^{\infty}(0,1), 
        \textrm{ as } n \to \infty, 
\end{equation}
where $\mathbb{G}$ is the standard Brownian bridge.
The limiting process is Gaussian with mean zero and covariance function 
\begin{equation*}
    \beta'_k \left( F^{-1}(u) \right) \beta'_k \left( F^{-1}(v) \right) \frac{ u \land v -uv }{f \left( F^{-1}(u) \right) f \left( F^{-1}(v) \right) }, \ u,v \in \R .
\end{equation*}
\end{lemma}

\begin{proof}[Proof of \cref{lemma19:asyNormal}]
Let $D[s,t]$ denote the Skorohod space of all cadlag functions on the interval $[s,t] \subset \bar{\R}$ with uniform norm.
Let $C[s,t]$ denote the space of all continuous functions on $[s,t]$.
Let $\ell^{\infty}(T)$ denote the collection of all bounded functions $f: T \mapsto \R$ with the norm $\normbig{f}_{\infty}=\sup_{t\in T}f$.

Let $\mathbb{D}_\phi $ be the set of all distribution functions with measures concentrated on $[s,t]$.
\Citet[Lemma 3.9.23(ii)]{vanderVaartWellner1996} show that
the inverse map $G \mapsto G^{-1}$ as a map 
$\phi: \mathbb{D}_\phi \subset D[s,t] \mapsto \ell^{\infty}(0,1) $ is Hadamard-differentiable at $F$ 
tangentially to $C[s,t]$.
The derivative is 
\begin{equation*}
    \phi'_{F} \left( \alpha \right)=- \left( \alpha/f \right) \circ F^{-1}=-\frac{\alpha\circ F^{-1} }{f \left( F^{-1} \right) } ,
\end{equation*}
for any function $\alpha \in D[s,t]$.

Consider the map 
$\psi: \ell^{\infty}(0,1) \mapsto \ell^{\infty}(0,1) $
given by 
$\psi \left( A \right) (x)=\beta_k \left( A(x) \right)=\beta_k \circ A(x) $. 
The domain of this map is the set of elements of $\ell^{\infty}(0,1)$ 
that take their values in the domain of $\beta_k$.
Let $\mathbb{D}_\psi=
\{ A: A\in \ell^{\infty}(0,1): s<A<t \}$. 
By \citet[Lemma 3.9.25]{vanderVaartWellner1996},
we know the map $A \mapsto \beta_k \circ A $ as a map 
$\mathbb{D}_\psi \subset \ell^{\infty}(0,1) \mapsto \ell^{\infty} \left( 0,1 \right) $
is Hadamard-differentiable at every $A \in \mathbb{D}_\psi$.
The derivative is $\psi'_A \left( \alpha \right)=\beta'_k \left( A(x) \right) \alpha(x) $.

Therefore, the map $F^{-1} \mapsto \beta_k \circ F^{-1}  $
as a map $\psi: \mathbb{D}_\psi \subset \ell^{\infty}(0,1) \mapsto \ell^{\infty}(0,1) $ is Hadamard-differentiable at every $F^{-1} \in \mathbb{D}_\psi$. 
The derivative is 
$\psi'_{F^{-1}} \left( \alpha \right) =\beta'_k \left( F^{-1}(x) \right) \alpha 
\left( x \right) $, for function $\alpha \in \ell^{\infty} \left( 0,1 \right) $.

By the chain rule, (See \citet[Lemma 3.9.3]{vanderVaartWellner1996}.)
the map  
$ \psi \circ \phi : \mathbb{D}_{\phi} \mapsto \ell^{\infty}(0,1) $ is Hadamard-differentiable at $F$
tangentially to $C[s,t]$, 
with derivative 
\begin{equation*}
    \left(\psi'_{\phi(F)} \circ \phi'_{F}
\right) (\alpha)=
\left(\psi'_{F^{-1}} \circ \phi'_{F}\right) (\alpha) 
= 
\beta'_k \left( F^{-1} \right) \phi'_F \left( \alpha \right)
=
\beta'_k \left( F^{-1} \right) \left( - \left( \alpha/f \right) \circ F^{-1} \right) ,
\end{equation*}
for any function $\alpha \in D[s,t]$.
That is, at any $p \in (0,1)$, the derivative is 
\begin{equation*}
    \left(\psi'_{\phi(F)} \circ \phi'_{F}
\right) \left( \alpha \right) \left( p \right) 
=\beta'_k \left( F^{-1} \left( p \right) \right) 
\left(- \frac{\alpha \left( F^
{-1} \left( p \right) \right) }{f \left( F^{-1} \left( p \right) \right) } \right) .
\end{equation*}

Applying the functional delta-methods 
(See \citet[Theorem 3.9.4]{vanderVaartWellner1996})
to the Donsker's theorem 
and the above result that the map $ \psi \circ \phi : \mathbb{D}_{\phi} \mapsto \ell^{\infty}(0,1) $ is Hadamard-differentiable at $F$
tangentially to $C[s,t]$,
\begin{align}
  \sqrt{n} \left( \beta_k \circ F_n^{-1} - \beta_k \circ F^{-1}
    \right)
    & = \sqrt{n} \left( \left(\psi \circ \phi \right) \left( F_n \right) - \left( \psi \circ \phi \right) \left( F \right)
    \right) \nonumber \\
    & \weaklyto 
        (\psi \circ \phi)'_F \left( \mathbb{G} \circ F \right) 
        \nonumber \\
     & = -\beta'_k \left( F^{-1} \right) \frac{ \left(\mathbb{G}\circ F \right) \circ F^{-1} }{f \left( F^{-1} \right) }
        \nonumber \\
    & =
    -\beta'_k \left( F^{-1} \right) \frac{ \mathbb{G} }{f \left( F^{-1} \right) }
        \nonumber \\
    & =
    \beta'_k \left( F^{-1} \right) \frac{ \mathbb{G} }{f \left( F^{-1} \right) },
     \ \textrm{in} \ \ell^{\infty}(0,1) ,
\end{align}
where $\mathbb{G}$ is the standard Brownian bridge. 
The last equality is due to the symmetry of the Gaussian process $\mathbb{G} $. 
The limiting process is Gaussian with mean zero and covariance function 
\begin{equation*}
    \beta'_k \left( F^{-1}(u) \right) 
    \beta'_k \left( F^{-1}(v) \right) 
    \frac{ u \land v -uv }{ f \left( F^{-1}(u) \right) f \left( F^{-1}(v) \right) }, \ u,v \in \R .
    \qedhere
\end{equation*}
\end{proof}

\section{Additional simulation results}\label{sec:appx:sim}

This section provides the additional simulation results tables.

\subsection{Additional simulation results for \texorpdfstring{\cref{sec:4.2:sim:varyXstar}}{Section \ref{sec:4.1:sim:varynT} }}\label{sec:B.2}

\cref{tab4.2:sim:varyXstar:n1000T1000:shift4:Vsd0.1,tab4.2:sim:varyXstar:n1000T10:shift4:Vsd0.1} present the results parallel to \cref{tab4.2:sim:varyXstar:n100T100:shift4:Vsd0.1} with a larger sample size $(n,T)=(1000,1000), (1000,10)$.
\cref{tab4.2:sim:varyXstar:n1000T1000:shift0} presents the simulation results in the case $X$ is unshifted, so that half of the $X$ takes negative value and may affect the sample order of $\hat{Y}^{*}$. I consider various sorting points $x^{*}=0, 0.5, 1, 1.5, 2, 2.5$. 
The results indicate that the choice of sorting points $x^{*}$ does not affect the sample estimates, as long as the population orders of the $Y^{*}$ are identical across these $x^{*}$.

\begin{table}[htbp]
    \centering\caption{\label{tab4.2:sim:varyXstar:n1000T1000:shift4:Vsd0.1}  MSE and Bias of $\hat{\beta}_1(\tau) $ } 
\sisetup{round-precision=3,round-mode=places}
    \begin{threeparttable}
    \begin{tabular}{llcS[table-format=-1.3]S[table-format=1.3]cS[table-format=-1.3]S[table-format=1.3]r@{}}
    \toprule
         &     &     & \multicolumn{2}{c}{ $\rho=0$} & &  \multicolumn{2}{c}{ $\rho=1$} \\           
    \cmidrule{4-5} \cmidrule{7-8}
         &     &     & {Bias} &  {MSE} &  & {Bias} & {MSE} &  \\      
    \cmidrule{4-8} 
$\hat{\beta}_1(0.25)$ & $x^{*}=   2.5$ & &  -0.000515  &  0.0000660  &  &   -0.000397 &  0.0000652   &    \\
                       & $x^{*}=   3.5$ & &  -0.000547  &  0.0000552  &  &   -0.000548 &  0.0000602   &    \\
                       & $x^{*}=   4.5$ & &  -0.000576  &  0.0000572  &  &   -0.000533 &  0.0000571   &    \\
                       & $x^{*}=   5.5$ & &  -0.000427  &  0.0000523  &  &   -0.000525 &  0.0000549   &    \\
                       & $x^{*}=   6.5$ & &  -0.000554  &  0.0000528  &  &   -0.000539 &  0.0000515   &    \\
\cmidrule{4-8} 
$\hat{\beta}_1(0.50)$ & $x^{*}=   2.5$ & &  -0.001316  &  0.0002734  &  &   -0.001029 &  0.0002851   &    \\
                      & $x^{*}=   3.5$ & &  -0.001006  &  0.0002608  &  &   -0.001128 &  0.0002643   &    \\ 
                      & $x^{*}=   4.5$ & &  -0.000993  &  0.0002636  &  &   -0.001047 &  0.0002610   &    \\ 
                      & $x^{*}=   5.5$ & &  -0.000772  &  0.0002640  &  &   -0.000659 &  0.0002683   &    \\ 
                      & $x^{*}=   6.5$ & &  -0.000816  &  0.0002537  &  &   -0.000627 &  0.0002578   &    \\ 
\cmidrule{4-8} 
$\hat{\beta}_1(0.75)$ & $x^{*}=   2.5$ & &  -0.000305  &  0.0004723  &  &   -0.000050 &  0.0004588   &    \\ 
                      & $x^{*}=   3.5$ & &  -0.000346  &  0.0004573  &  &   -0.000620 &  0.0004653   &    \\
                      & $x^{*}=   4.5$ & &  -0.000447  &  0.0004456  &  &   -0.000296 &  0.0004562   &    \\
                      & $x^{*}=   5.5$ & &  -0.000487  &  0.0004458  &  &   -0.000518 &  0.0004448   &    \\
                      & $x^{*}=   6.5$ & &  -0.000401  &  0.0004473  &  &   -0.000528 &  0.0004463   &    \\
\bottomrule
            \end{tabular}
            \begin{tablenotes}
            \item $500$ replications. $(n, T)=(1000, 1000)$. Shift 4. $\sigma_V=0.1$.
            \end{tablenotes}
            \end{threeparttable}
            \end{table}

\begin{table}[htbp]
    \centering\caption{\label{tab4.2:sim:varyXstar:n1000T10:shift4:Vsd0.1}  MSE and Bias of $\hat{\beta}_1(\tau) $ } 
\sisetup{round-precision=3,round-mode=places}
    \begin{threeparttable}
    \begin{tabular}{llcS[table-format=-1.3]S[table-format=1.3]cS[table-format=-1.3]S[table-format=1.3]r}
    \toprule
         &     &     & \multicolumn{2}{c}{ $\rho=0$} & &  \multicolumn{2}{c}{ $\rho=1$} \\           
    \cmidrule{4-5} \cmidrule{7-8}
         &     &     & {Bias} &  {MSE} &  & {Bias} & {MSE} &  \\      
    \cmidrule{4-8} 
$\hat{\beta}_1(0.25)$ & $x^{*}=   2.5$ & &  0.0008512  &  0.0025676  &  &   -0.003192 &  0.0025801   &    \\
                       & $x^{*}=   3.5$ & &  0.0010842  &  0.0017160  &  &   -0.000136 &  0.0018398   &    \\
                       & $x^{*}=   4.5$ & &  0.0004211  &  0.0012688  &  &   -0.000261 &  0.0013497   &    \\
                       & $x^{*}=   5.5$ & &  0.0013023  &  0.0009584  &  &   -0.000649 &  0.0009483   &    \\
                       & $x^{*}=   6.5$ & &  0.0022886  &  0.0007118  &  &   0.0023358 &  0.0009946   &    \\
\cmidrule{4-8} 
$\hat{\beta}_1(0.50)$ & $x^{*}=   2.5$ & &  0.0005713  &  0.0031625  &  &   -0.004885 &  0.0040687   &    \\
                      & $x^{*}=   3.5$ & &  -0.000820  &  0.0023850  &  &   -0.001281 &  0.0027252   &    \\ 
                      & $x^{*}=   4.5$ & &  -0.000479  &  0.0015259  &  &   0.0020421 &  0.0018395   &    \\ 
                      & $x^{*}=   5.5$ & &  -0.001251  &  0.0011211  &  &   0.0026322 &  0.0011672   &    \\ 
                      & $x^{*}=   6.5$ & &  -0.000498  &  0.0010207  &  &   0.0011731 &  0.0009529   &    \\ 
\cmidrule{4-8} 
$\hat{\beta}_1(0.75)$ & $x^{*}=   2.5$ & &  0.0004397  &  0.0036109  &  &   -0.002347 &  0.0048609   &    \\ 
                      & $x^{*}=   3.5$ & &  -0.004289  &  0.0019480  &  &   0.0048269 &  0.0028987   &    \\
                      & $x^{*}=   4.5$ & &  0.0001426  &  0.0016768  &  &   -0.003850 &  0.0020919   &    \\
                      & $x^{*}=   5.5$ & &  -0.001445  &  0.0012209  &  &   0.0000969 &  0.0016211   &    \\
                      & $x^{*}=   6.5$ & &  -0.000205  &  0.0011936  &  &   0.0007742 &  0.0012390   &    \\
\bottomrule
            \end{tabular}
            \begin{tablenotes}
            \item $500$ replications. $(n, T)=(1000, 10)$. Shift 4. $\sigma_V=0.1$.
            \end{tablenotes}
            \end{threeparttable}
            \end{table}

\begin{table}[htbp]
    \centering\caption{\label{tab4.2:sim:varyXstar:n1000T1000:shift0}  MSE and Bias of $\hat{\beta}_1(\tau) $ } 
\sisetup{round-precision=3,round-mode=places}
    \begin{threeparttable}
    \begin{tabular}{llcS[table-format=-1.3]S[table-format=1.3]cS[table-format=-1.3]S[table-format=1.3]r@{}}
    \toprule
         &     &     & \multicolumn{2}{c}{ $\rho=0$} & &  \multicolumn{2}{c}{ $\rho=1$} \\           
    \cmidrule{4-5} \cmidrule{7-8}
         &     &     & {Bias} &  {MSE} &  & {Bias} & {MSE} &  \\      
    \cmidrule{4-8} 
$\hat{\beta}_1(0.25)$ & $x^{*}=     0$ & &  -0.000682  &  0.0000616  &  &   -0.000645 &  0.0000600   &    \\
                       & $x^{*}=   0.5$ & &  -0.000415  &  0.0000548  &  &   -0.000272 &  0.0000543   &    \\
                       & $x^{*}=     1$ & &  -0.000652  &  0.0000529  &  &   -0.000568 &  0.0000525   &    \\
                       & $x^{*}=   1.5$ & &  -0.000497  &  0.0000528  &  &   -0.000544 &  0.0000511   &    \\
                       & $x^{*}=     2$ & &  -0.000545  &  0.0000510  &  &   -0.000568 &  0.0000513   &    \\
                       & $x^{*}=   2.5$ & &  -0.000572  &  0.0000493  &  &   -0.000541 &  0.0000503   &    \\
\cmidrule{4-8} 
$\hat{\beta}_1(0.50)$ & $x^{*}=     0$ & &  -0.000926  &  0.0002641  &  &   -0.001170 &  0.0003008   &    \\
                      & $x^{*}=   0.5$ & &  -0.000887  &  0.0002589  &  &   -0.000814 &  0.0002650   &    \\ 
                      & $x^{*}=     1$ & &  -0.000847  &  0.0002611  &  &   -0.000923 &  0.0002608   &    \\ 
                      & $x^{*}=   1.5$ & &  -0.000827  &  0.0002556  &  &   -0.000783 &  0.0002576   &    \\ 
                      & $x^{*}=     2$ & &  -0.000851  &  0.0002544  &  &   -0.000800 &  0.0002562   &    \\ 
                      & $x^{*}=   2.5$ & &  -0.000869  &  0.0002554  &  &   -0.000938 &  0.0002540   &    \\ 
\cmidrule{4-8} 
$\hat{\beta}_1(0.75)$ & $x^{*}=     0$ & &  -0.000769  &  0.0004810  &  &   -0.000695 &  0.0005364   &    \\ 
                      & $x^{*}=   0.5$ & &  -0.000161  &  0.0004507  &  &   -0.000512 &  0.0004588   &    \\
                      & $x^{*}=     1$ & &  -0.000463  &  0.0004532  &  &   -0.000538 &  0.0004479   &    \\
                      & $x^{*}=   1.5$ & &  -0.000428  &  0.0004521  &  &   -0.000345 &  0.0004430   &    \\
                      & $x^{*}=     2$ & &  -0.000399  &  0.0004463  &  &   -0.000333 &  0.0004441   &    \\
                      & $x^{*}=   2.5$ & &  -0.000410  &  0.0004450  &  &   -0.000329 &  0.0004451   &    \\
\bottomrule
            \end{tabular}
            \begin{tablenotes}
            \item $500$ replications. $(n, T)=(1000, 1000)$. Shift 0. $\sigma_V=0.1$.
            \end{tablenotes}
            \end{threeparttable}
            \end{table}

\subsection{Additional simulation results for \texorpdfstring{\cref{sec:4.4:sim:FEQR}}{Section \ref{sec:4.1:sim:varynT} }}\label{sec:B.4}

\Cref{tab4.4:sim:n1000:T10} is the parallel version of \cref{tab4.4:sim:n100:T100}
with $(n,T)=(1000,10)$.

\begin{table}[htbp]
     \fontsize{11}{12}\selectfont
    \centering\caption{\label{tab4.4:sim:n1000:T10}  MSE and Bias of $\hat{\beta}_1(\tau) $ } 
\sisetup{round-precision=3,round-mode=places}
\begin{adjustbox}{width=0.9\columnwidth,center}
    \begin{threeparttable}
    \begin{tabular}{llccS[table-format=-1.3]S[table-format=1.3]cS[table-format=-1.3]S[table-format=1.3]c}
    \toprule
 $\tilde{V}_{it} \sim N(0, \sigma_V^2)$     &               &                        &     & \multicolumn{2}{c}{My model ($x^{*}=4$)   }   & & \multicolumn{2}{c}{FE-QR \citep{Canay2011}   }       &      \\
    \cmidrule{5-6} \cmidrule{8-9} 
    &               &                        &     &    {Bias}    &       {MSE}        & &     {Bias}     &       {MSE}         &  \\      
     
\cmidrule{5-9} 
$\sigma_V=0.01$ & $\rho=  0$   & $\hat{\beta}_1(0.25)$  &     &  -0.000136  &  0.0001867  & &   0.2215447 &  0.0492442   &    \\
&               & $\hat{\beta}_1(0.50)$  &     &  -0.000411  &  0.0005435  & &   0.0196452 &  0.0005682   &    \\
&               & $\hat{\beta}_1(0.75)$  &     &  -0.000060  &  0.0011107  & &   -0.277441 &  0.0771440   &    \\
\cmidrule{5-9} 
& $\rho=  1$    & $\hat{\beta}_1(0.25)$  &     &  -0.000180  &  0.0002041  & &   0.2053457 &  0.0423619   &    \\
&               & $\hat{\beta}_1(0.50)$  &     &  -0.001047  &  0.0007590  & &   0.0238539 &  0.0007392   &    \\
&               & $\hat{\beta}_1(0.75)$  &     &  -0.001700  &  0.0016692  & &   -0.262311 &  0.0689454   &    \\
\cmidrule{5-9} 
& $\rho=  3$    & $\hat{\beta}_1(0.25)$  &     &  -0.000097  &  0.0002460  & &   0.2139595 &  0.0459900   &    \\
&               & $\hat{\beta}_1(0.50)$  &     &  0.0009330  &  0.0012796  & &   0.0476370 &  0.0024058   &    \\
&               & $\hat{\beta}_1(0.75)$  &     &  0.0018762  &  0.0040803  & &   -0.244183 &  0.0597384   &    \\
\cmidrule{5-9} 
& $\rho= 10$    & $\hat{\beta}_1(0.25)$  &     &  0.0040175  &  0.0008570  & &   0.2560862 &  0.0657029   &    \\
&               & $\hat{\beta}_1(0.50)$  &     &  0.0159530  &  0.0084780  & &   0.0754338 &  0.0057998   &    \\
&               & $\hat{\beta}_1(0.75)$  &     &  0.0235540  &  0.0323256  & &   -0.228879 &  0.0524881   &    \\
\cmidrule{2-9} 
$\sigma_V=0.1$ & $\rho=  0$   & $\hat{\beta}_1(0.25)$  &     &  0.0055410  &  0.0136888  & &   0.1932276 &  0.0374681   &    \\
&               & $\hat{\beta}_1(0.50)$  &     &  0.0079321  &  0.0335847  & &   -0.015046 &  0.0003921   &    \\
&               & $\hat{\beta}_1(0.75)$  &     &  0.0035951  &  0.0572132  & &   -0.239044 &  0.0573352   &    \\
\cmidrule{5-9} 
& $\rho=  1$    & $\hat{\beta}_1(0.25)$  &     &  0.0176103  &  0.0157766  & &   0.1304176 &  0.0171494   &    \\
&               & $\hat{\beta}_1(0.50)$  &     &  0.0196063  &  0.0527173  & &   -0.020827 &  0.0005894   &    \\
&               & $\hat{\beta}_1(0.75)$  &     &  0.0291931  &  0.0946263  & &   -0.209451 &  0.0440482   &    \\
\cmidrule{5-9} 
& $\rho=  3$    & $\hat{\beta}_1(0.25)$  &     &  0.0584137  &  0.0566750  & &   0.0667918 &  0.0045905   &    \\
&               & $\hat{\beta}_1(0.50)$  &     &  0.1107193  &  0.1343058  & &   0.0102767 &  0.0002742   &    \\
&               & $\hat{\beta}_1(0.75)$  &     &  0.0198576  &  0.1752389  & &   -0.167995 &  0.0283875   &    \\
\cmidrule{5-9} 
& $\rho= 10$    & $\hat{\beta}_1(0.25)$  &     &  0.0448207  &  0.0441919  & &   0.1315785 &  0.0175485   &    \\
&               & $\hat{\beta}_1(0.50)$  &     &  -0.005701  &  0.1234348  & &   0.0683439 &  0.0049196   &    \\
&               & $\hat{\beta}_1(0.75)$  &     &  -0.230545  &  0.2208454  & &   -0.135937 &  0.0187074   &    \\
\cmidrule{2-9} 
$\sigma_V=1$ & $\rho=  0$   & $\hat{\beta}_1(0.25)$  &     &  0.1789398  &  0.2950929  & &   0.0530406 &  0.0030142   &    \\
&               & $\hat{\beta}_1(0.50)$  &     &  0.0749575  &  0.3127882  & &   -0.000915 &  0.0003962   &    \\
&               & $\hat{\beta}_1(0.75)$  &     &  -0.098572  &  0.3526222  & &   -0.031946 &  0.0016116   &    \\
\cmidrule{5-9} 
& $\rho=  1$    & $\hat{\beta}_1(0.25)$  &     &  0.3526257  &  0.4582600  & &   0.0149896 &  0.0004648   &    \\
&               & $\hat{\beta}_1(0.50)$  &     &  0.1792945  &  0.3440336  & &   0.0200372 &  0.0008694   &    \\
&               & $\hat{\beta}_1(0.75)$  &     &  -0.223461  &  0.4015186  & &   0.0068743 &  0.0007097   &    \\
\cmidrule{5-9} 
& $\rho=  3$    & $\hat{\beta}_1(0.25)$  &     &  0.3845798  &  0.3976957  & &   -0.026827 &  0.0010509   &    \\
&               & $\hat{\beta}_1(0.50)$  &     &  0.0601258  &  0.2456467  & &   0.0554691 &  0.0036111   &    \\
&               & $\hat{\beta}_1(0.75)$  &     &  -0.494590  &  0.4703020  & &   0.0586611 &  0.0041130   &    \\
\cmidrule{5-9} 
& $\rho= 10$    & $\hat{\beta}_1(0.25)$  &     &  0.8821320  &  0.8742174  & &   -0.016200 &  0.0012030   &    \\
&               & $\hat{\beta}_1(0.50)$  &     &  -0.004282  &  0.1579474  & &   0.0709121 &  0.0060561   &    \\
&               & $\hat{\beta}_1(0.75)$  &     &  -0.820041  &  0.8287285  & &   0.0639256 &  0.0051296   &    \\
\bottomrule
            \end{tabular}
            \begin{tablenotes}
            \item $500$ replications. $(n, T)=(1000, 10)$. Shift 4.
            \end{tablenotes}
    \end{threeparttable}
    \end{adjustbox}
\end{table}

\subsection{Compare with standard FE}\label{sec:4.3:sim:stdFE}

\begin{table}[htbp]
    \centering
    \caption{\label{tab8:sim:compareFE:n100:T100}  MSE and Bias of $\hat{\beta}_1(\tau) $ } 
\sisetup{round-precision=3,round-mode=places}
    \begin{threeparttable}
    \begin{tabular}{llcS[table-format=-1.3]S[table-format=1.3]cS[table-format=-1.3]S[table-format=1.3]cS[table-format=-1.3]S[table-format=1.3]cS[table-format=-1.3]S[table-format=1.3]r@{}}
    \toprule
         &     &     & \multicolumn{2}{c}{ $\rho=0$} & &  \multicolumn{2}{c}{ $\rho=1$}  & &  \multicolumn{2}{c}{ $\rho=3$}  & &  \multicolumn{2}{c}{ $\rho=10$} \\           
    \cmidrule{4-5} \cmidrule{7-8}  \cmidrule{10-11}  \cmidrule{13-14}
         &     &     & {Bias} &  {MSE} &  & {Bias} & {MSE} &  & {Bias} & {MSE} &  & {Bias} & {MSE} &  \\      
    \cmidrule{4-14} 
$\hat{\beta}_1(0.25)$ & $x^{*}=     5$ & &  0.0003618  &  0.0005471  &  &   0.0004908 &  0.0005508   &   &   0.0003678 &  0.0005407   &  &   -0.000271 &  0.0004970   &   \\
                       & $x^{*}=     6$ & &  0.0001642  &  0.0005385  &  &   0.0000612 &  0.0005257   &   &   0.0000555 &  0.0005247   &  &   -0.000056 &  0.0004969   &   \\
                       & $x^{*}=     7$ & &  -0.000065  &  0.0005329  &  &   0.0000522 &  0.0005169   &   &   0.0000852 &  0.0005186   &  &   0.0002914 &  0.0005034   &   \\
                       & $x^{*}=     8$ & &  0.0000328  &  0.0005329  &  &   0.0003342 &  0.0005084   &   &   0.0001998 &  0.0005031   &  &   0.0002249 &  0.0005024   &   \\
\cmidrule{4-14} 
$\hat{\beta}_1(0.50)$ & $x^{*}=     5$ & &  -0.005782  &  0.0024612  &  &   -0.005133 &  0.0024001   &   &   -0.005269 &  0.0024862   &  &   -0.005078 &  0.0024923   &   \\
                      & $x^{*}=     6$ & &  -0.005328  &  0.0024139  &  &   -0.005782 &  0.0023957   &    &   -0.005541 &  0.0024741   &  &   -0.005163 &  0.0024664   &  \\ 
                      & $x^{*}=     7$ & &  -0.005619  &  0.0024094  &  &   -0.005967 &  0.0024110   &    &   -0.005208 &  0.0024205   &  &   -0.005422 &  0.0024645   &  \\ 
                      & $x^{*}=     8$ & &  -0.005425  &  0.0024169  &  &   -0.005560 &  0.0024075   &    &   -0.005455 &  0.0024036   &  &   -0.005635 &  0.0024566   &  \\ 
\cmidrule{4-14} 
$\hat{\beta}_1(0.75)$ & $x^{*}=     5$ & &  -0.012280  &  0.0043770  &  &   -0.011673 &  0.0042730   &   &   -0.012030 &  0.0043501   &  &   -0.011975 &  0.0043630   &   \\ 
                      & $x^{*}=     6$ & &  -0.012233  &  0.0043360  &  &   -0.011656 &  0.0042791   &   &   -0.011908 &  0.0043252   &  &   -0.012111 &  0.0043576   &   \\
                      & $x^{*}=     7$ & &  -0.012325  &  0.0043447  &  &   -0.011948 &  0.0043055   &   &   -0.011705 &  0.0042843   &  &   -0.011978 &  0.0043199   &   \\
                      & $x^{*}=     8$ & &  -0.012160  &  0.0043219  &  &   -0.012155 &  0.0042986   &   &   -0.011711 &  0.0042726   &  &   -0.011782 &  0.0043063   &   \\
\cmidrule{4-14} 
$\hat{\beta}_{1, \mathrm{FE}}$ &                & &  -0.002139  &  0.0009275  &  &   0.1059404 &  0.0123542   &   &   0.1815426 &  0.0341167   &  &   0.2337705 &  0.0557800   &   \\ 
\bottomrule
            \end{tabular}
            \begin{tablenotes}
            \item $500$ replications. $(n,T)=(100,100)$. $\sigma_V=0.1$.
            \end{tablenotes}
    \end{threeparttable}
\end{table}

First, the standard FE model cannot estimate heterogeneous causal effect. 
My model extends the standard FE model to allow heterogeneous causal effect.

Second, the standard FE model can have additional problems of endogeneity bias as shown in the following simulation.
Consider the same DGP as in \cref{eqn:sim:4.1} except that the regressor $X_{it}= (1+\rho U_i)\times (X_{ind, it}+4)$. 
Like in \cref{eqn:sim:4.1}, the $X_{ind,it}$ is shifted to right by 4 standard deviation; the $\rho$ measures the (nonadditive) endogeneity level of the DGP.
The majority of $X_{it}$ take values on [5,8]. So I choose sorting points $x^{*}=5,6,7,8$.
The slope coefficient is $\beta_1(U)=U^2$ with $U\sim \UnifDist(0,1)$. I consider $\E(\beta_1(U))=\E(U^2)=1/3$ as the true parameter value to compare with standard FE estimator.

\Cref{tab8:sim:compareFE:n100:T100} reports the bias and MSE of my estimator and the standard FE estimator.
It shows when the model has endogeneity ($\rho=1, 3, 10$), the standard FE estimator starts to have large bias, and thus large MSE. 
And the bias and MSE of FE estimator increases as the endogeneity level increases. When $\rho=10$, the FE estimator has 0.23 bias and 0.056 MSE.
My estimator has a low bias (up to 0.012) and low MSE (up to 0.004) with all endogeneity levels at all quantile levels $\tau=0.25, 0.5, 0.75$.
The same as found in \cref{sec:4.2:sim:varyXstar}, the sorting points do not matter much for my estimator, as long as population orders of $Y^{*}$ are identical across the sorting points.

\section{Additional details for empirical example}\label{sec:appx:emp}

\subsection{Bootstrap algorithm for standard error}
\label{appx:bs-se-algorithm}

\Cref{algorithm:bs-se} presents the algorithm for the bootstrap estimator of standard error. 
It is computed following \citep[p.2222--2223]{ChernozhukovEtAl2013DR}.
\Citet[Remark 3.2 and Lemma SA.1]{ChernozhukovEtAl2013DR} 
have discussed and proved the validity of the bootstrap standard error computed via \Cref{algorithm:bs-se}. 
Replication code is available online at \url{https://xinliu16.github.io/} for additional details.

\begin{algorithm}
\caption{(Bootstrap algorithm for standard error)}\label{algorithm:bs-se}
  Let $n$ denote the sample size on the cross-sectional dimension. Let $\matf{W}_i=(\vecf{Y}_i, \matf{X}_i)$ denote the iid (over $i$) sampling from \cref{A1:iidSampling}. 
  \begin{enumerate}[label=\arabic*., ref=\arabic*]
    \item \label{bs:sample} Draw the bootstrap sample $\matf{W}_{i}^{*(b)}$ from the original sample ($\matf{W}_1,\ldots,\matf{W}_n$) randomly with replacement.  
    \item \label{bs:stat} Compute the estimator 
    $\hat{\vecf{\beta}}{}^{(b)}(\tau)$ 
    from the bootstrap sample $\matf{W}_{i}^{*(b)}$.
    \item Repeat steps \cref{bs:sample,bs:stat} for $b=1,\ldots,B$, where $B$ is the number of bootstraps.
    \item Compute the bootstrap standard error as $\mathrm{SE}(\hat{\beta}_\tau)=(q^{*}_{0.75}-q^{*}_{0.25})/(z_{0.75}-z_{0.25})$, 
    where $q^{*}_{p}$ is the sample $p$th  quantile of among $\hat{\vecf{\beta}}^{(b)}(\tau)$ over $b=1, \ldots, B$ and $z_p$ is the $p$th quantile of $N(0,1)$.
\end{enumerate}
\end{algorithm}

\subsection{Additional empirical results with other \texorpdfstring{$x^{*}$}{x*} values }\label{appx:emp:x-star}

This section presents additional empirical results illustrating the robustness of estimates locally around the sorting points $\vecf{x}^*$ considered in \cref{sec:emp}.
Specifically, I let the sorting point values for the variable of interest $\log( \textit{OILWEALTH})$ vary from 1.1 to 2 with a general increment of 0.1 and with a finer grid around its sample mean. 
The variable of interest $\log( \textit{OILWEALTH})$ takes value ranging from $-2.7734$ to $8.2681$ with a sample mean of $1.8229$, which is the sorting point value used in \cref{tab:empirical}.
The sorting point values for other covariates remain the same as used in \cref{tab:empirical}.

\Cref{tab:emp:x-star} reports the estimates for the same parameter of interest as in \cref{tab:empirical} at rank values $\tau=\{0.25, 0.5, 0.75\}$ using these additional sorting point values. 
All of the three estimates ($\hat{\beta}(0.25)$, $\hat{\beta}(0.5)$, and $\hat{\beta}(0.75)$) remain the same across the sorting point values in the range (1.3, 1.84). 
The $\hat{\beta}(0.5)$ changes to 0.25 when the sorting point value decreases to 1.2, while the $\hat{\beta}(0.25)$ and $\hat{\beta}(0.75)$ are not affected by this sorting point change. 
The $\hat{\beta}(0.25)$ changes to $-1.11$ when the sorting point value increases to 2.
The $\hat{\beta}(0.75)$ changes to $-0.13$ when the sorting point value increases to 1.85. 
These evidences indicate the robustness of estimates locally around the sorting point $x^*$ used in \cref{tab:empirical}.

Although \cref{tab:emp:x-star} indicates the monotonicity assumed in \cref{A3:monoto} is plausible locally (especially within the (1.3,1.85) sorting value interval), 
we cannot rule out the possibility that the monotonicity might still hold over a wider region of sorting points values even if we observe the changes in estimates.
Indeed, the estimated rank order (among the units) may not catch the true latent population rank order due to the finite sample estimation error.
In other words, the order-discovering is an asymptotic result in \cref{prop4:betanorm-nrate}.

It is interesting to derive a formal test for checking whether the monotonicity holds simultaneously at at least two points as well as quantify the sampling uncertainty to identify and estimate the monotonicity region. 
I leave these extensions for future research.

\begin{table}[t]
        \centering\caption{\label{tab:emp:x-star} Estimates for the effect of oil wealth on military defense spending using various sorting points.}
\sisetup{round-precision=2,round-mode=places}
                \begin{threeparttable}
                \begin{tabular}{lcS[table-format=-1.2]S[table-format=1.2]
                cS[table-format=-1.2]S[table-format=1.2]
                cS[table-format=-1.2]S[table-format=1.2]r@{}}
                \toprule
              &     & \multicolumn{2}{c}{ $\tau=0.25$} & &  \multicolumn{2}{c}{ $\tau=0.5$}  & &  \multicolumn{2}{c}{ $\tau=0.75$}   \\           
                \cmidrule{3-4}     \cmidrule{6-7}  \cmidrule{9-10}
                &     & {$\hat{\beta}_\tau$}     &  {$\mathrm{SE}(\hat{\beta}_\tau)$} &     
                      & {$\hat{\beta}_\tau$}     &  {$\mathrm{SE}(\hat{\beta}_\tau)$} &    
                      & {$\hat{\beta}_\tau$}     &  {$\mathrm{SE}(\hat{\beta}_\tau)$} &   \\
                \cmidrule{3-10}
 $x^*=$       1.1   & &       -0.339   &       0.2254  & &      0.2493   &      0.2473  & &      -0.394   &      0.2299  &   \\
 $x^*=$       1.2   & &       -0.339   &       0.2954  & &      0.2493   &      0.2868  & &      -0.394   &      0.2077  &   \\
 $x^*=$       1.3   & &       -0.339   &       0.2429  & &      -0.176   &      0.3044  & &      -0.394   &      0.2052  &   \\
 $x^*=$       1.4   & &       -0.339   &       0.2954  & &      -0.176   &      0.3044  & &      -0.394   &      0.2058  &   \\
 $x^*=$       1.5   & &       -0.339   &       0.2850  & &      -0.176   &      0.2473  & &      -0.394   &      0.2052  &   \\
 $x^*=$       1.6   & &       -0.339   &       0.2954  & &      -0.176   &      0.2770  & &      -0.394   &      0.1868  &   \\
 $x^*=$       1.7   & &       -0.339   &       0.2816  & &      -0.176   &      0.2473  & &      -0.394   &      0.2114  &   \\
 $x^*=$       1.8   & &       -0.339   &       0.3590  & &      -0.176   &      0.2771  & &      -0.394   &      0.2324  &   \\
 $x^*=$    1.8229   & &       -0.339   &       0.3590  & &      -0.176   &      0.2473  & &      -0.394   &      0.2299  &   \\
 $x^*=$      1.84   & &       -0.339   &       0.3057  & &      -0.176   &      0.3044  & &      -0.394   &      0.2114  &   \\
 $x^*=$      1.85   & &       -0.339   &       0.2954  & &      -0.176   &      0.3044  & &      -0.129   &      0.2299  &   \\
 $x^*=$       1.9   & &       -0.339   &       0.2679  & &      -0.176   &      0.3044  & &      -0.129   &      0.2082  &   \\
 $x^*=$         2   & &       -1.113   &       0.2954  & &      -0.176   &      0.3173  & &      -0.129   &      0.2299  &   \\
\bottomrule
                \end{tabular}
                \begin{tablenotes}
                \item  100 bootstrap replications. $x^*=1.8229$ is the sorting point value used in \cref{tab:empirical}.
                \end{tablenotes}
                \end{threeparttable}
                \end{table}

\end{appendices}

\end{document}